\documentclass[aps,prc,twocolumn,superscriptaddress]{revtex4-2}
\usepackage{graphicx}% Include figure files
\usepackage{dcolumn}% Align table columns on decimal point
\usepackage{textgreek}
\begin{document}
	
	% Use the \preprint command to place your local institutional report
	% number in the upper righthand corner of the title page in preprint mode.
	% Multiple \preprint commands are allowed.
	% Use the 'preprintnumbers' class option to override journal defaults
	% to display numbers if necessary
	%\preprint{}
	
	%Title of paper
	\title{Neutron Induced Fission Fragment Angular Distributions, Anisotropy, and Linear Momentum Transfer Measured with the NIFFTE Fission Time Projection Chamber}
	
	% repeat the \author .. \affiliation  etc. as needed
	% \email, \thanks, \homepage, \altaffiliation all apply to the current
	% author. Explanatory text should go in the []'s, actual e-mail
	% address or url should go in the {}'s for \email and \homepage.
	% Please use the appropriate macro foreach each type of information
	
	% \affiliation command applies to all authors since the last
	% \affiliation command. The \affiliation command should follow the
	% other information
	% \affiliation can be followed by \email, \homepage, \thanks as well.
	
	%\email[]{Your e-mail address}
	%\homepage[]{Your web page}
	%\thanks{}
	%\altaffiliation{}
	\author{D. Hensle}
	\email{dhensle@mines.edu}
	\affiliation{Colorado School of Mines, Golden, Colorado 80401, USA}
	
	\author{ J.T. Barker}	\affiliation{Abilene Christian University, Abilene, Texas 79699}
	\author{ J.S. Barrett}	\affiliation{Oregon State University, Corvallis, Oregon 97331}
	\author{ N.S. Bowden}	\affiliation{Lawrence Livermore National Laboratory, Livermore, California 94550}
	\author{ K.J. Brewster}	\affiliation{Abilene Christian University, Abilene, Texas 79699}
	\author{ J. Bundgaard}	\affiliation{Colorado School of Mines, Golden, Colorado 80401}
	\author{ Z.Q. Case}	\affiliation{Abilene Christian University, Abilene, Texas 79699}
	\author{ R.J. Casperson}	\affiliation{Lawrence Livermore National Laboratory, Livermore, California 94550}
	\author{ D.A. Cebra}	\affiliation{University of California, Davis, California 95616}
	\author{ T. Classen}	\affiliation{Lawrence Livermore National Laboratory, Livermore, California 94550}
	\author{ D.L. Duke}	\affiliation{Los Alamos National Laboratory, Los Alamos, New Mexico 87545}
	\author{ N. Fotiadis}	\affiliation{Los Alamos National Laboratory, Los Alamos, New Mexico 87545}
	\author{ J Gearhart}	\affiliation{University of California, Davis, California 95616}
	\author{ V. Geppert-Kleinrath}	\affiliation{Los Alamos National Laboratory, Los Alamos, New Mexico 87545}
	\author{ U. Greife}	\affiliation{Colorado School of Mines, Golden, Colorado 80401}
	\author{ E. Guardincerri}	\affiliation{Los Alamos National Laboratory, Los Alamos, New Mexico 87545}
	\author{ C. Hagmann}	\affiliation{Lawrence Livermore National Laboratory, Livermore, California 94550}
	\author{ M. Heffner}	\affiliation{Lawrence Livermore National Laboratory, Livermore, California 94550}
	\author{C.R. Hicks}	\affiliation{Abilene Christian University, Abilene, Texas 79699}
	\author{ D. Higgins}	\affiliation{Los Alamos National Laboratory, Los Alamos, New Mexico 87545}
	\author{ L.D. Isenhower}	\affiliation{Abilene Christian University, Abilene, Texas 79699}
	\author{ K. Kazkaz}	\affiliation{Lawrence Livermore National Laboratory, Livermore, California 94550}
	\author{ A. Kemnitz}	\affiliation{California Polytechnic State University, San Luis Obispo, California 93407}
	\author{ K.J. Kiesling}	\affiliation{Abilene Christian University, Abilene, Texas 79699}
	\author{ J. King}	\affiliation{Oregon State University, Corvallis, Oregon 97331}
	\author{ J.L. Klay}	\affiliation{California Polytechnic State University, San Luis Obispo, California 93407}
	\author{ J. Latta}	\affiliation{Colorado School of Mines, Golden, Colorado 80401}
	\author{ E. Leal}	\affiliation{Los Alamos National Laboratory, Los Alamos, New Mexico 87545}
	\author{ W. Loveland}	\affiliation{Oregon State University, Corvallis, Oregon 97331}
	\author{ M. Lynch}	\affiliation{California Polytechnic State University, San Luis Obispo, California 93407}
	\author{ J.A. Magee} \altaffiliation[Currently at ]{Stonehill College}	\affiliation{Lawrence Livermore National Laboratory, Livermore, California 94550} 
	\author{ B. Manning}	\affiliation{Los Alamos National Laboratory, Los Alamos, New Mexico 87545}
	\author{ M.P. Mendenhall}	\affiliation{Lawrence Livermore National Laboratory, Livermore, California 94550}
	\author{ M. Monterial}	\affiliation{Lawrence Livermore National Laboratory, Livermore, California 94550}
	\author{ S. Mosby}	\affiliation{Los Alamos National Laboratory, Los Alamos, New Mexico 87545}
	\author{ G. Oman}	\affiliation{California Polytechnic State University, San Luis Obispo, California 93407}
	\author{ C. Prokop}	\affiliation{Los Alamos National Laboratory, Los Alamos, New Mexico 87545}
	\author{ S. Sangiorgio}	\affiliation{Lawrence Livermore National Laboratory, Livermore, California 94550}
	\author{ K.T. Schmitt}	\affiliation{Los Alamos National Laboratory, Los Alamos, New Mexico 87545}
	\author{ B. Seilhan}	\affiliation{Lawrence Livermore National Laboratory, Livermore, California 94550}
	\author{ L. Snyder}	\affiliation{Lawrence Livermore National Laboratory, Livermore, California 94550}
	\author{ F. Tovesson}	\affiliation{Los Alamos National Laboratory, Los Alamos, New Mexico 87545}
	\author{ C.L. Towell}	\affiliation{Abilene Christian University, Abilene, Texas 79699}
	\author{ R.S. Towell}	\affiliation{Abilene Christian University, Abilene, Texas 79699}
	\author{ T.R. Towell}	\affiliation{Abilene Christian University, Abilene, Texas 79699}
	\author{ N. Walsh}	\affiliation{Lawrence Livermore National Laboratory, Livermore, California 94550}
	\author{ T.S. Watson}	\affiliation{Abilene Christian University, Abilene, Texas 79699}
	\author{ L. Yao}	\affiliation{Oregon State University, Corvallis, Oregon 97331}
	\author{ W. Younes}	\affiliation{Lawrence Livermore National Laboratory, Livermore, California 94550}

	%Collaboration name if desired (requires use of superscriptaddress
	%option in \documentclass). \noaffiliation is required (may also be
	%used with the \author command).
	%\collaboration can be followed by \email, \homepage, \thanks as well.
	\collaboration{NIFFTE Collaboration} \homepage{http://niffte.calpoly.edu/}
	%\noaffiliation
	
	\date{\today}

	\begin{abstract}
		The Neutron Induced Fission Fragment Tracking Experiment (NIFFTE) collaboration has performed measurements with a fission time projection chamber (fissionTPC) to study the fission process by reconstructing full three-dimensional tracks of fission fragments and other ionizing radiation. The amount of linear momentum imparted to the fissioning nucleus by the incident neutron can be inferred by measuring the opening angle between the fission fragments.  Using this measured linear momentum, fission fragment angular distributions can be converted to the center-of-mass frame for anisotropy measurements.  Angular anisotropy is an important experimental observable for understanding the quantum mechanical state of the fissioning nucleus and vital to determining detection efficiency for cross section measurements. Neutron linear momentum transfer to fissioning $^{235}$U, $^{238}$U, and $^{239}$Pu and fission fragment angular anisotropy of $^{235}$U and $^{238}$U as a function of neutron energies in the range 130\,keV--250\,MeV are presented.
	\end{abstract}
	
	% insert suggested keywords - APS authors don't need to do this
	%\keywords{}
	
	%\maketitle must follow title, authors, abstract, and keywords
	\maketitle
	
	% body of paper here - Use proper section commands
	% References should be done using the \cite, \ref, and \label commands
	\section{Introduction}
	Nuclear fission data is an important input into many applications including nuclear reactors, stockpile stewardship, and astrophysics \cite{Aliberti2006}. Neutron induced fission cross sections are of particular importance, but are difficult to measure with the precision required for some applications \cite{Salvatores2007}.  Thus, the Neutron Induced Fission Fragment Tracking Experiment (NIFFTE) collaboration has built a time projection chamber, the fissionTPC, designed specifically to study the fission process \cite{Heffner2014}.  The fissionTPC has the ability to perform three-dimensional reconstruction of ionizing radiation in the detector volume.  By leveraging this three-dimensional tracking ability, the fissionTPC has the capability to explore cross section systematics not accessible by other fission detection methods \cite{Casperson2018}.  Concurrently with cross section measurements, fission fragment angular distributions and linear momentum transfer from the incident neutron to the target nucleus are measured.
	
	Bohr first introduced the idea of a fissioning nucleus being thermodynamically cold due to all of the energy being stored in the deformation of the nucleus \cite{Bohr1956}, allowing the fissioning nucleus to be described by a discrete angular momentum state.  Fission fragment angular distributions can be calculated from analytical expressions based on the angular momentum state of the fissioning nucleus \cite{VandenboschText,Lamphere1962}.  At high excitation energies, these discrete angular momentum states are expected to turn into a continuous regime \cite{Soheyli2012,Behkami1968}.  Experimentally, a sum of different angular momentum states contributes to the measured angular distribution of fission fragments at each incident neutron energy.
	
	Fission fragment angular distributions can be used to infer the angular momentum state of the fissioning nucleus. These results provide useful empirical inputs to Monte Carlo fission reaction codes such as TALYS \cite{talys} and EMPIRE \cite{empire}.  Conversely, comparison of the measured angular distributions with the outputs from these codes provides a benchmark for the included in reaction models.  Angular anisotropy is also included in cross section measurements to fully understand detector efficiency \cite{Casperson2018}. 
	
	Due to the linear momentum transfer of the incident neutron to the target nucleus, the emission angles of fission fragments in the center-of-mass frame are focused forward in the lab frame.  Fission fragments also gain additional kinetic energy in the direction of the neutron flight path.  Consequently, more fission events are directed downstream of the target and fewer fission fragments are detected in the upstream volume, thus affecting the detector efficiency as a function of neutron energy in any cross section measurement \cite{Casperson2018}.  Other fission observables such as fragment energies and masses must also take this kinematic boost into account \cite{Higgins2018}.
	
	In this paper, we present measurements of fission fragment angular anisotropy for $^{235}$U and $^{238}$U as well as mean neutron linear momentum transfer to fissioning $^{235}$U, $^{238}$U, and $^{239}$Pu in the neutron energy range 130 keV -- 250 MeV.
	
	\section{Background}
	Fission fragment angular distributions are fit with Legendre polynomials of even order to conserve forward-backward symmetry \cite{VandenboschText}.  An anisotropy parameter is then reported to display this information as a function of incident neutron energies and is typically defined as the ratio of counts parallel to the beam to counts perpendicular to the beam.  This anisotropy parameter is expressed as 
	\begin{equation}
	\label{eqn:A}
	A = \frac{W[\cos\theta_{c.m.} = 1]}{W[\cos\theta_{c.m.} = 0]} = \frac{W(0^{\circ}_{c.m.})}{W(90^{\circ}_{c.m.})}
	\end{equation}
	where, in the case of this work,
	\begin{equation}
	W(\cos\theta_{c.m.}) = a_{0} + a_{2} L_{2}( \cos\theta_{c.m.}) + a_{4} L_{4}(\cos\theta_{c.m.})
	\end{equation}
	and $\theta_{c.m.}$ is the fission fragment polar angle in the center-of-mass frame.  
	
	Because the anisotropy parameter is defined in the center-of-mass frame, a conversion from the lab frame to the center-of-mass frame is needed.  Past experiments have accounted for this correction by averaging upstream and downstream results \cite{Vorobyev2016}, assuming full momentum transfer \cite{Verena}, or treating it as a negligible correction \cite{Tarrio2014}.  A necessary input parameter to make this correction without these approximations is the amount of linear momentum the fissioning nucleus has actually acquired from the incident neutron, but, to the best of our knowledge, no such measurement for neutron-induced fission exists.
	
	Above neutron energies of a few MeV, new reaction mechanisms arise that complicate the simple picture of complete neutron capture followed by fission.  Direct reactions and pre-equilibrium particle emission \cite{HodgsonText} become intermediate steps between incident neutron interaction and scission.  These reaction mechanisms result in light particles (primarily neutrons and protons) being emitted in the direction of the neutron beam \cite{Bertrand1973} and necessarily take away some of the available linear momentum carried by the incident neutron.  By measuring the opening angle of the fission fragments, the amount of linear momentum transfer occurring as a function of incident neutron energy can be extracted and used to convert fission fragments from the lab frame to the center of mass frame -- a method first proposed by Sikkeland \textit{et al.}  \cite{Sikkeland1962} and expanded upon in this paper.

	\section{Detector}
\begin{figure}[]
		\includegraphics[width=.5\textwidth]{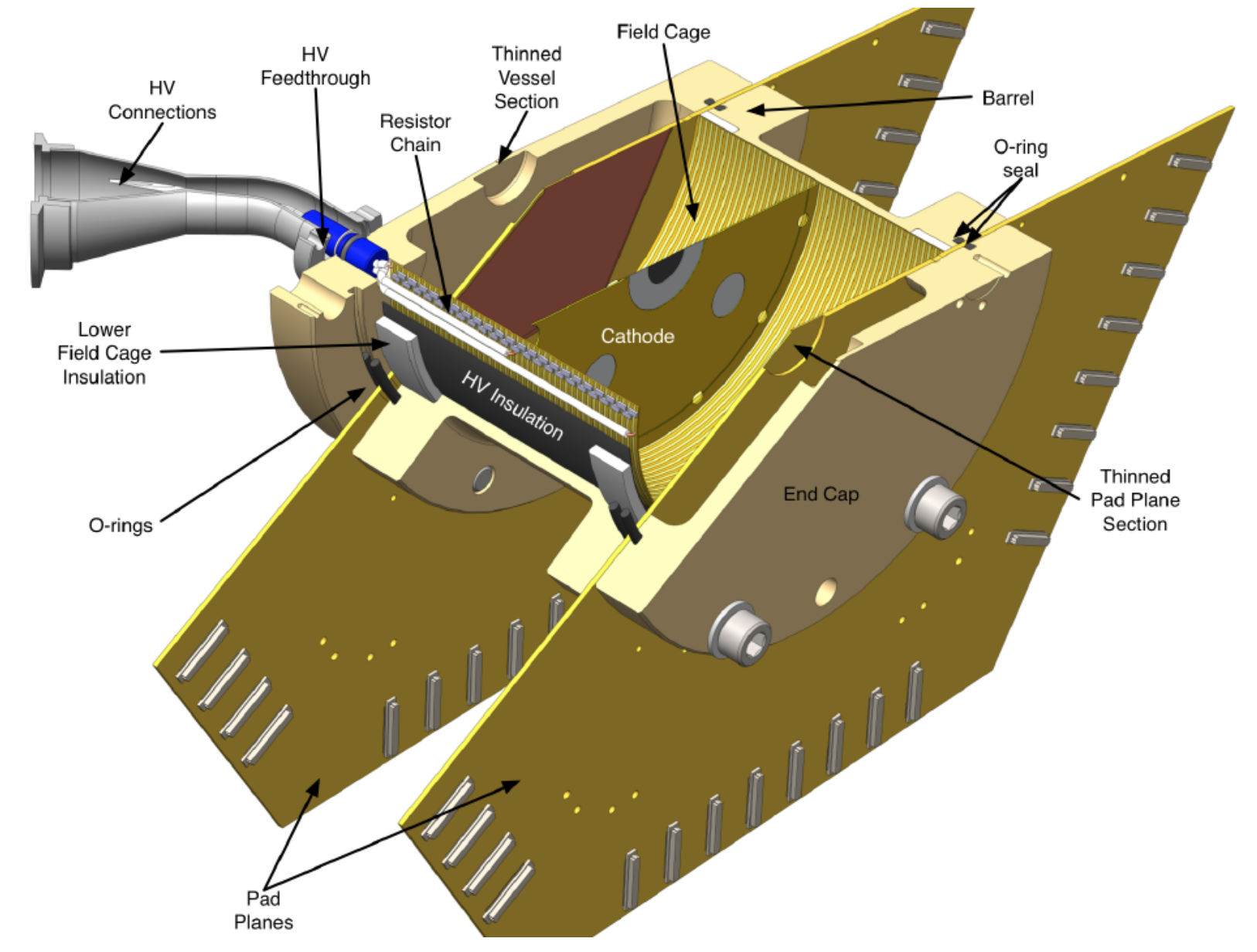}
		\caption{ Schematic drawing of the fissionTPC hardware showing two gas volumes, each with a highly segmented anode, separated by a central cathode.  The actinide targets are placed on the central cathode.}
		\label{fig:tpchardware}
	\end{figure}

	The fissionTPC, shown in Figure \ref{fig:tpchardware} and described in detail in \cite{Heffner2014}, consists of two gas volumes separated by a central cathode. Ionizing radiation strips electrons from the fill gas which are drifted to the anode pad planes by a static electric field of 500\,V/cm.  Each volume has a highly segmented anode with 2976 hexagonal readout pads, 2\,mm in pitch, and a MICROMEGAS (MICRO MEsh Gaseous Structure) metal mesh \cite{Giomataris1996} held 75\,\textmu m above the pads for gas amplification to produce a gain factor of approximately 40 \cite{Snyder2018}.  The drift chamber, filled with 95\% argon / 5\% isobutane at 550\,torr \cite{Snyder2018}, is 15\,cm in diameter and each anode is 5.4\,cm from the central cathode.  Each of the anode pads are connected to custom electronic readout cards with a sampling rate of 20\,ns. The central cathode is used for neutron time of flight and has a sampling rate of one nanosecond \cite{Heffner2013}.
	
\begin{figure}[]
		\includegraphics[width=.5\textwidth]{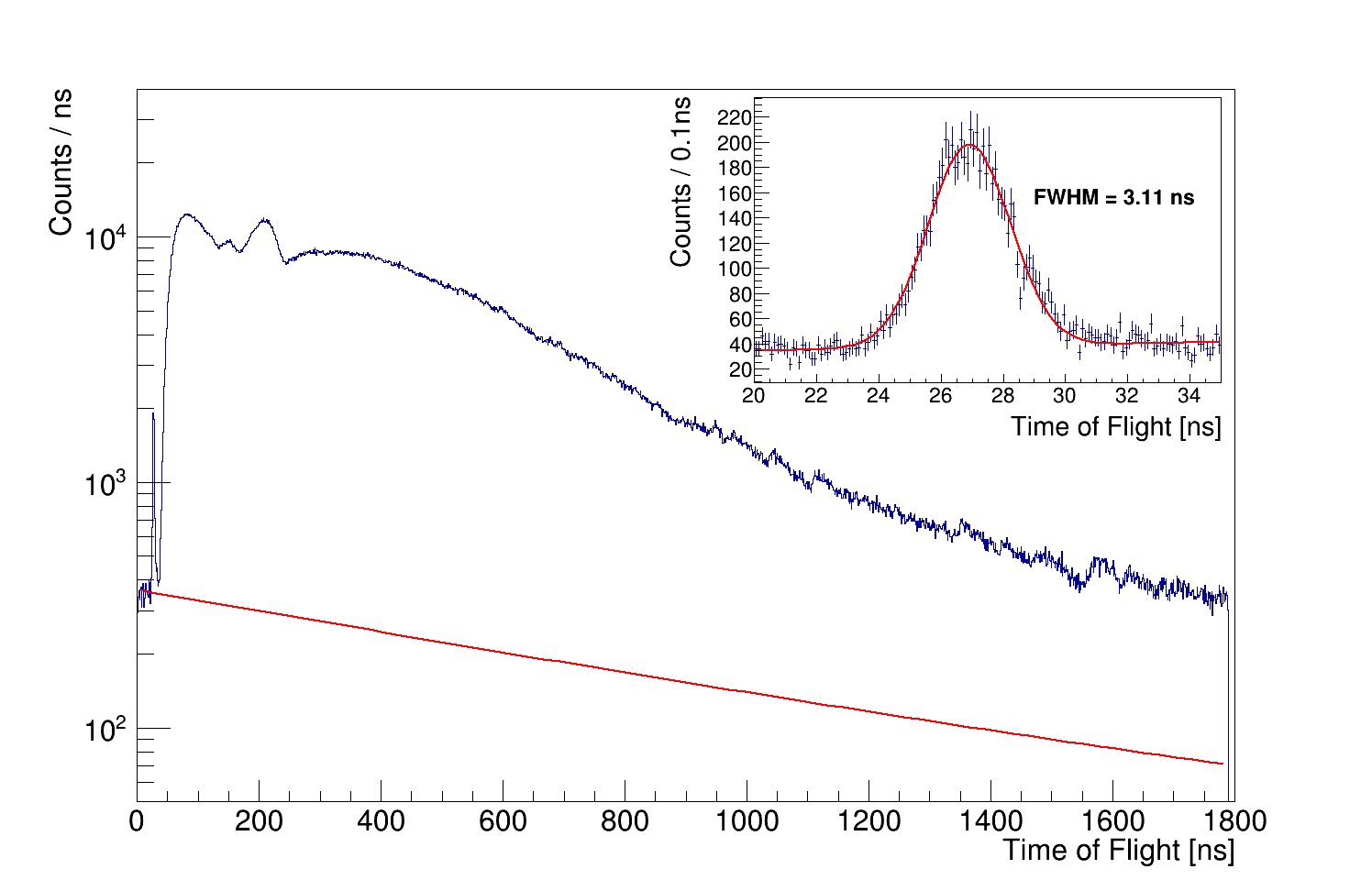}
		\caption{Neutron time-of-flight spectrum for a $^{235}$U target where the red line represents the amount of neutron beam wraparound.  The inset shows the peak at 26\,ns resulting from photon induced fission and has a FWHM of about 3\,ns, demonstrating the time-of-flight resolution of the fissionTPC.}
		\label{fig:nToF}
	\end{figure}

	The fissionTPC operates at the Los Alamos National Laboratory's Neutron Science Center at the Weapons Neutron Research flight path 90L.  An 800\,MeV proton beam impinges upon a tungsten target to create an unmoderated white neutron flux.  The structure of the neutron beam is typically set to a 100\,Hz macropulse structure containing about 375 micropulses of 250\,ps in width, each spaced by 1.8\,\textmu s \cite{Lisowski1990}.  Neutron time of flight is used to extract the neutron kinetic energy.  The start signal is the proton pulse hitting the spallation target and the stop signal is fragment detection by the central cathode. Figure \ref{fig:nToF} shows the neutron time of flight spectrum for a $^{235}$U target. The peak at $\sim$26\,ns is due to photon induced fission and is used to calibrate the time of flight spectrum as well as shows the time-of-flight resolution of approximately 3\,ns FWHM.
	
	Due to the neutron beam micropulse separation of only 1.8\,\textmu s, neutrons from the subsequent pulse can pass the slowest neutrons from the previous pulse. These slow neutrons are then assigned the time of flight with respect to the most recent micropulse, not the micropulse from which they originated.  The amount of these wraparound neutrons can be estimated by fitting the tail of the last micropulse and summing that fit over all micropulses.  This procedure produces the red line in Figure \ref{fig:nToF} which must match the pedestal before the photo-fission peak as these neutrons must be coming from wraparound events considering no neutrons can arrive before the gamma flash.
	
	\section{Actinide Targets}
	
\begin{figure}[]
		\includegraphics[width=.47\textwidth]{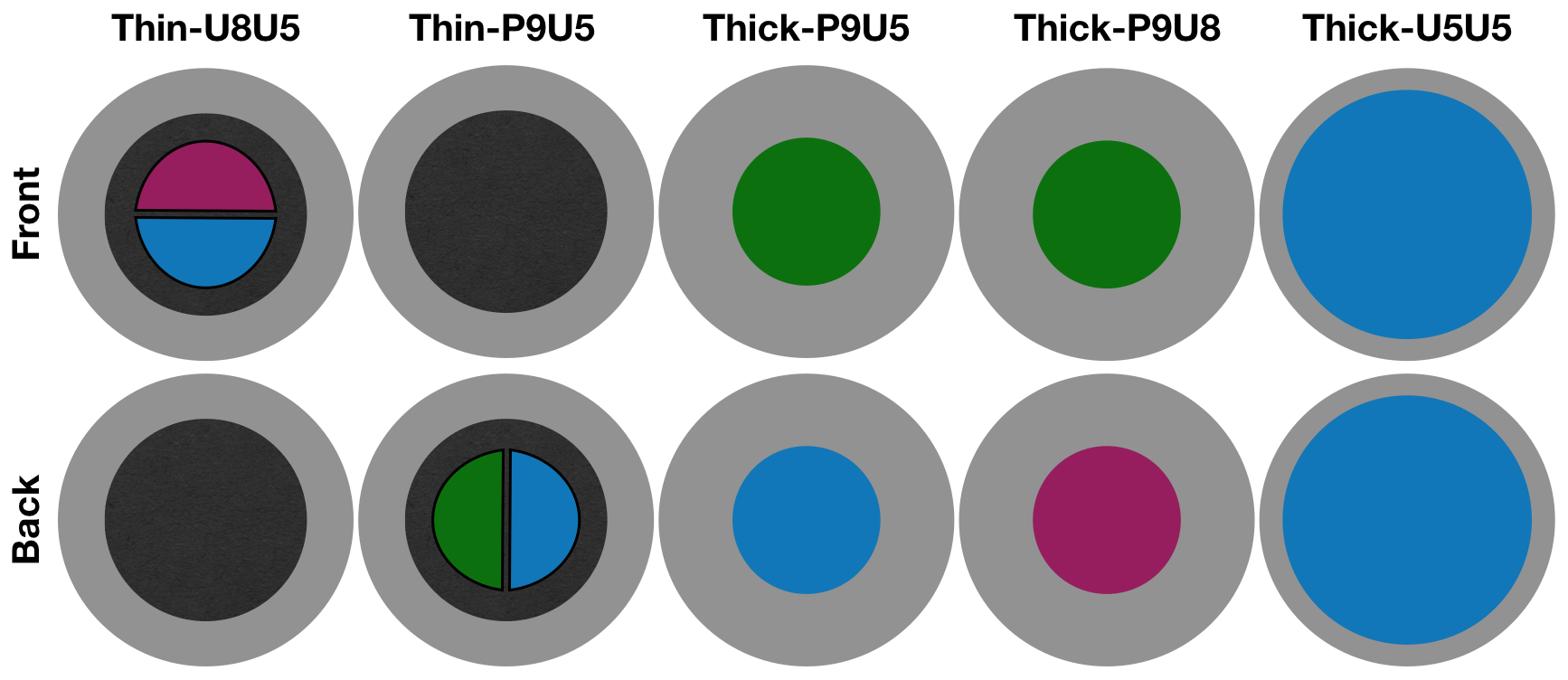}
		\caption{List of targets used in this analysis.  The first two targets are placed on a thin carbon backing whereas the other targets are on thick aluminum backings.}
		\label{fig:targets}
	\end{figure}

	Three actinides, $^{235}$U, $^{238}$U, and $^{239}$Pu, have been placed in the fissionTPC and exposed to the neutron source in a number of different target configurations, as shown in Figure \ref{fig:targets}.  The first two targets consist of half-moons of the actinides placed on 100\,\textmu g/cm$^{2}$ carbon, allowing fission fragments and alpha particles emitted from the actinide deposit to be detected in both volumes.  All of the other targets are placed on aluminum backings with a thickness of 0.5\,mm which effectively separates the upstream and downstream gas volumes as fission fragments and alpha particles cannot pass through the aluminum backing.  In each of the thick target runs, the fissionTPC was rotated with respect to the beam such that each side has the kinematic boost from the incident neutron flipped between the upstream and downstream data sets.  Changing targets requires the entire fissionTPC to be disassembled to gain access to the central cathode target holder.
	
	One additional target of $^{252}$Cf (not shown in Figure \ref{fig:targets}) was placed in the fissionTPC to calibrate the detector response to alpha particles and fission fragments without a neutron beam.  This source provides isotropic emission of fission fragments and alpha particles to demonstrate the ability of the fissionTPC to accurately measure a known angular distribution.
	
	\section{Data Reconstruction}
	A large amount of processing must be done on the raw fissionTPC data in order to extract ionizing track parameters.  Voxels of charge are created which contain the x, y, z locations, and charge in units of uncalibrated Analog to Digital Converter (ADC) units.  First, the anode pad signals undergo a differentiation process to extract the amount of charge that was incident on the pad at each individual time step.  The pad position gives the x and y values for each voxel, and the z information is extracted by multiplying the electron drift speed by the drift time relative to the start of the event.  Because the fissionTPC cannot measure exactly when ionization occurs within the chamber, the z values are relative and not absolute.
	
	%\begin{figure}[]
	%	\includegraphics[width=.5\textwidth]{figures/digits.png}
	%	\caption{A visual representation of two fission fragments and other ionizing events within the fissionTPC.  The red lines are the fits to each track and the color scale represents the amount of charge in each voxel, in uncalibrated ADC units.}
	%	\label{fig:digits}
	%\end{figure}
	
	\begin{figure}[]
		\includegraphics[width=.5\textwidth]{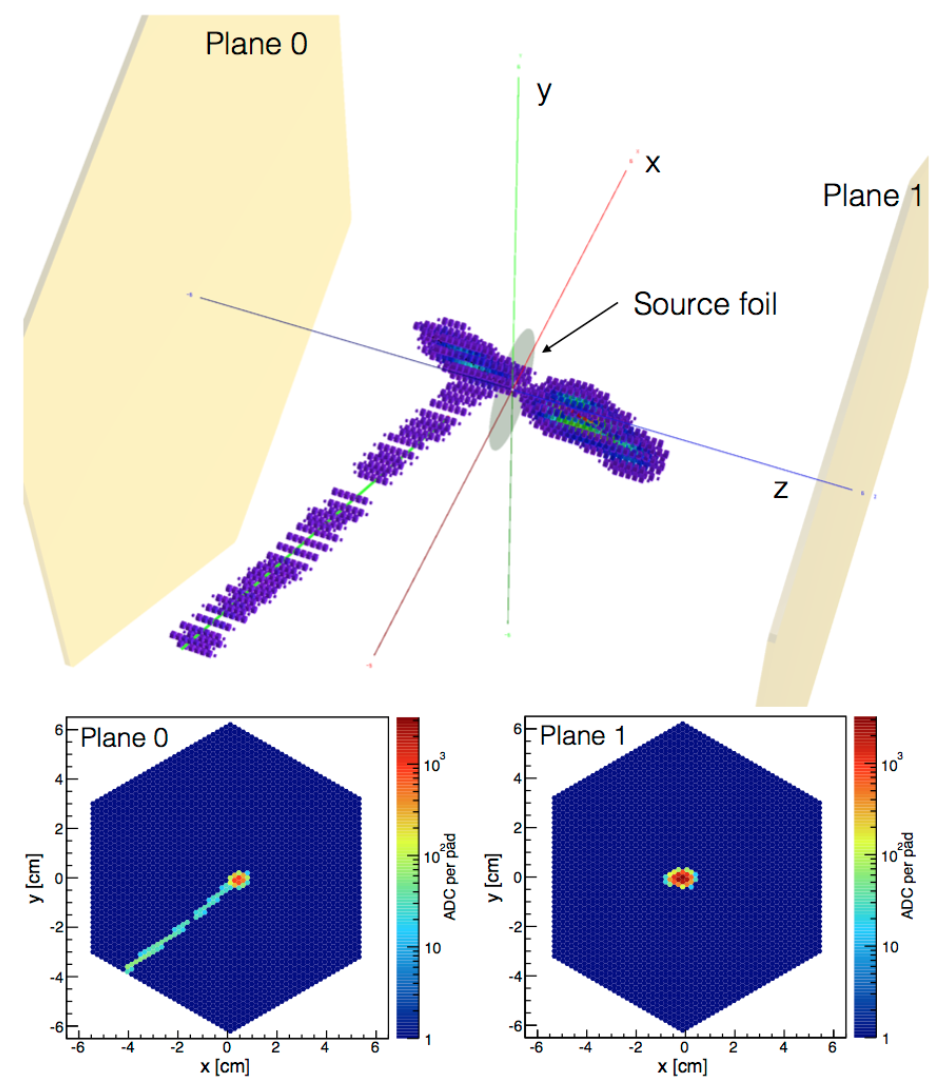}
		\caption{Visualization of reconstructed fission fragment and ternary alpha particle charge clouds in the fissionTPC from an actinide target with a thin carbon backing allowing both fission fragments to be detected.  The color scale represents the amount of charge in each voxel in uncalibrated ADC units \cite{Heffner2014}.}
		\label{fig:digits}
	\end{figure}
	
	Once all of the voxels are created, they are grouped together and fit with a straight line. Figure \ref{fig:digits} shows an example event from a thin-backed target containing the upstream and downstream fission fragments and a ternary alpha particle \cite{Heffner2014}.
	
\begin{figure}[]
		\includegraphics[width=.47\textwidth]{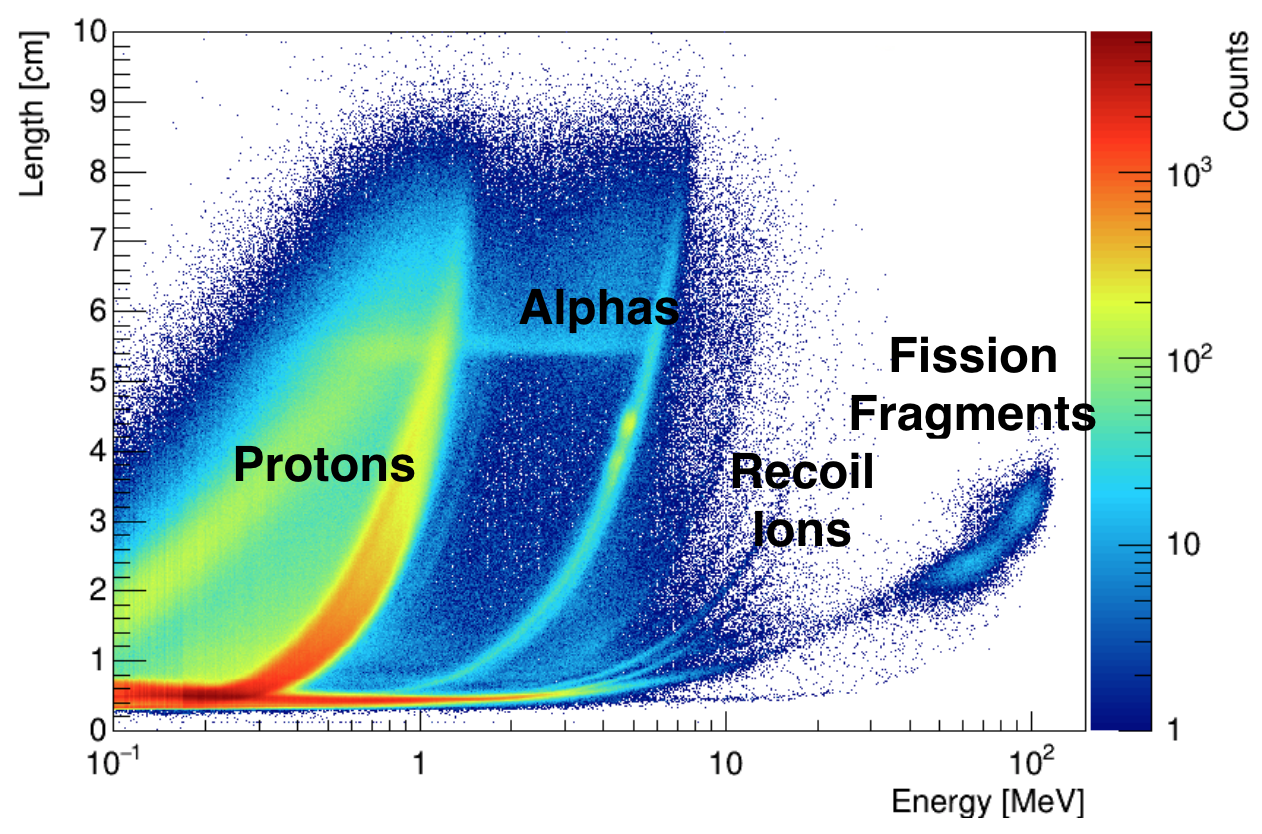}
		\caption{Track length vs total track energy provides a common parameter space which aids particle identification in the fissionTPC \cite{Casperson2018}.}
		\label{fig:lvadc}
	\end{figure}

	From each track fit, a number of track parameters can be extracted including the start and end positions, polar and azimuthal angles, peak ionization value and location, length, and charge.  In Figure \ref{fig:lvadc}, a plot of track length versus track energy shows how a common parameter space is used to differentiate types of ionizing radiation \cite{Casperson2018}.  Fission fragments are shorter and more energetic while alpha particles travel farther and deposit less energy.  Rough energy calibrations can be done using the known energy of alpha particles from spontaneous decays to convert from ADC to MeV.
	
	In the context of fission fragment angular anisotropy and linear momentum transfer measurements, the polar angle of each track is the primary observable.  However, corrections for electron drift speed and electron diffusion must be applied to each fission fragment track to accurately extract these polar angles from the data.
	
	\subsection{Electron Drift Speed}\label{sub:electronDrift}
	The fissionTPC does not have the capability to directly measure the speed at which the electrons drift through the chamber.  However, an accurate measurement of the drift speed is needed in order to properly reconstruct the polar angles of the ionizing radiation.  The cosine of the polar angle is determined by 
	\begin{equation}
	\cos\theta = \frac{\Delta z}{l}
	\end{equation}
	where $\Delta z$ is the track length in the drift direction of the chamber and $l$ is the total track length. Considering the z position of each charge voxel is determined by multiplying the relative drift time by the electron drift speed, a larger drift speed produces a larger $\Delta z$ and vice-versa. Thus, changing the electron drift speed changes the reconstructed track polar angle.
	
	In order to find the correct electron drift speed in the fissionTPC, the spontaneous alpha emission from the actinide target is used.  Because the spontaneous alpha emission must be isotropic, the $\cos\theta$ distribution must be flat.  Thus, the electron drift speed in the reconstruction process is adjusted to achieve a flat polar angle distribution.
	
	Using this technique, a drift speed for each target in each configuration is found with an associated uncertainty of 0.01\,cm/\textmu s.  Because electron drift speed is dependent on gas temperature and pressure, as well as the applied drift voltage, the drift speed measurement between the targets used in this analysis varies from 2.75 to 3.09\,cm/\textmu s \cite{hensleThesis}.
	
	\subsection{Electron Diffusion}\label{sub:electronDiff}
	After setting the electron drift speed to reproduce isotropic emission of alpha particles, an additional polar angle correction for electron diffusion is needed for the fission fragments.  Because the fission fragments have a much higher ionization density than the alpha particles, electron diffusion manifests more strongly in the fission fragments.  If the electron diffusion coefficients are different between the radial and drift directions of the fissionTPC, then a polar angle distortion takes place.
	
	By treating the electron diffusion as a Gaussian distribution with different widths in the radial and drift directions of the fissionTPC, a correction to the polar angle takes the form \cite{hensleThesis}
	\begin{equation}
	\label{eqn:electronDiffusion}
	\delta \cos\theta = \frac{\sin^{2}\theta \cos\theta}{\int a^{2}q(a)da / Q} <l> (\sigma_{r0}^{2} - \sigma_{z0}^{2})
	\end{equation}
	where $a$ is the length along the track with respect to the center of the charge cloud, $Q$ is the total charge in the track, $q(a)$ is the charge per unit length along the track, $\sigma_{r0,z0}$ are the electron diffusion coefficients in the radial and drift directions, respectively. The average electron drift length is given by 
	\begin{equation}
	<l> = \frac{1}{Q} \int l(a)q(a)da
	\end{equation}
	where $l(a) = 5.4\,\text{cm} - a \cdot \cos\theta$ and 5.4\,cm is the length of the fissionTPC in the drift direction.  Due to the large amount of data needed to save the complete Bragg curve for each fission fragment, $q(a)$ is estimated by treating the Bragg curve as having a Gaussian start and a linear tail where the two functions meet at the Bragg peak and the slope of the linear tail is such that $\int q(a)da = Q$.  Thus, since $q(a)$ is estimated using only three points -- start and end positions, and the Bragg peak -- the data requirements in the processing step are greatly reduced.
	
	Fission fragments emitted perpendicular to the neutron beam direction, i.e. with $0 < \cos\theta < 0.05$, have width projections along the drift direction or the radial direction of the fissionTPC.  By stacking many fragment charge clouds on top of each other, the standard deviations of the charge cloud projections, $\sigma_{r}$ and $\sigma_{z}$, can be calculated. These charge cloud widths are related to the diffusion coefficients by the drift distance they traveled
	\begin{equation}
	\sigma_{r,z}^{2} = \sigma_{r0,z0}^{2}<l>
	\end{equation}
	where $<l> = 5.4$\,cm for these fragments that are perpendicular to the neutron beam direction.
	
	Since electron diffusion coefficients are dependent on gas parameters, they are also measured for each target in each configuration with values of $\sigma_{z}$ ranging from 0.126 to  0.141\,cm and values of $\sigma_{r}$ ranging from 0.152 to 0.161\,cm with an uncertainty of 0.001\,cm.  This produces a 10\% uncertainty on the overall diffusion correction.  The diffusion correction uncertainty associated with modifying Bragg curve parameters was negligible when compared to the uncertainty caused by the diffusion coefficients \cite{hensleThesis}.
	
	\begin{figure}[]
		\includegraphics[width=.5\textwidth]{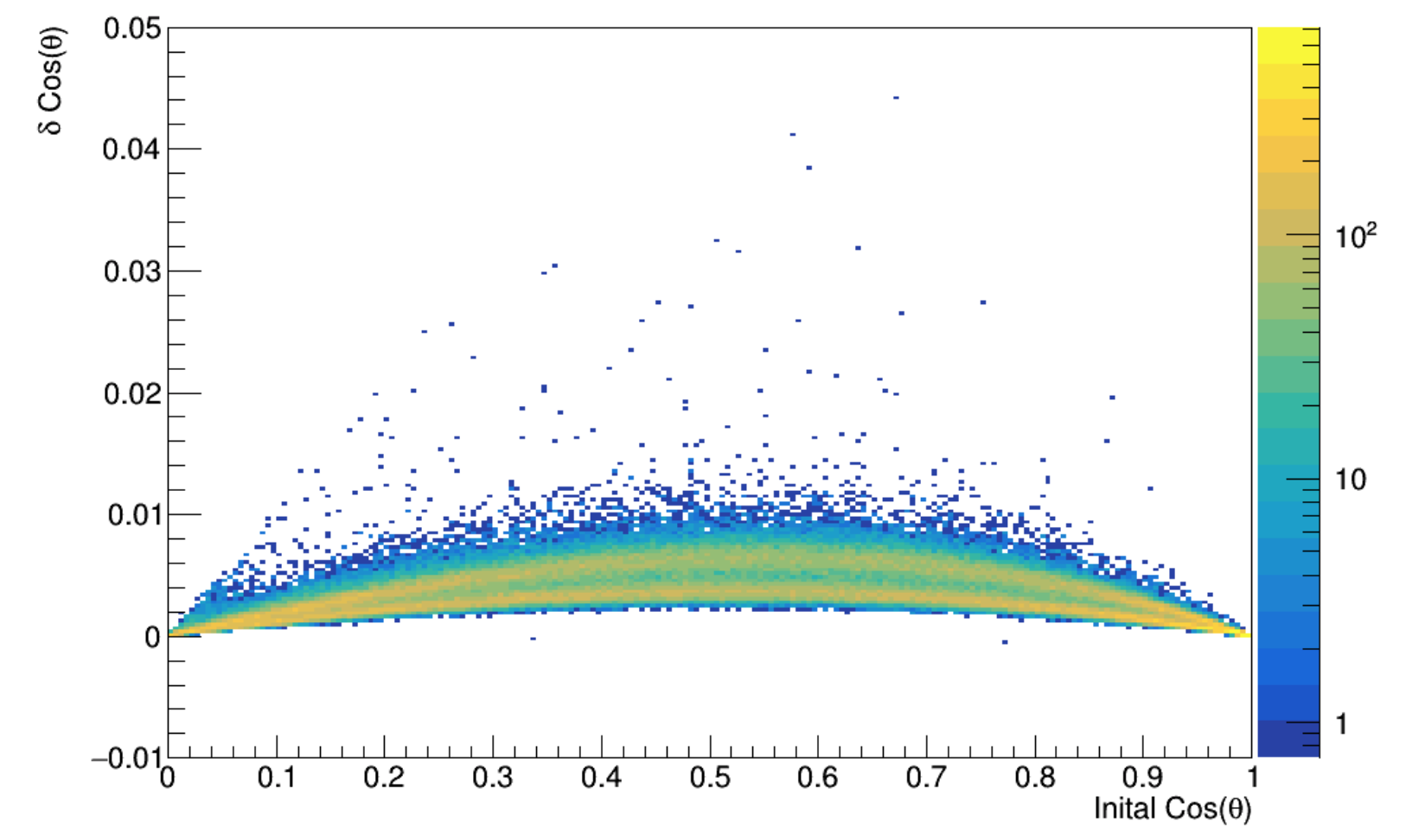}
		\caption{Values calculated by Eq. \ref{eqn:electronDiffusion} showing the magnitude of the electron diffusion correction for the $^{252}$Cf source.  The two bands are from the light and heavy fragments which have slightly different correction magnitudes due to their different ionization densities.}
		\label{fig:diffusion}
	\end{figure}
	
	\subsection{Californium-252 Calibration Source}\label{sub:cf252}
	
	To ensure the electron diffusion and electron drift speed corrections successfully reproduce fragment angular distributions, a source of $^{252}$Cf was placed in the fissionTPC to provide a simultaneous source of isotropic fission fragments and alpha particles. This measurement of the track widths resulted in $\sigma_{r} = 0.156$\,cm and $\sigma_{z} = 0.133$\,cm.  Notice that the closer $\sigma_{r}$ is to $\sigma_{z}$, the smaller the correction in Equation \ref{eqn:electronDiffusion}.  A completely symmetric electron diffusion would have $\sigma_{r} = \sigma_{z}$, thus leading to no correction.  The magnitude and shape of the correction can be seen in Figure \ref{fig:diffusion} where the two bands result from the light and heavy fragments.
	
\begin{figure}[]
		\includegraphics[width=.5\textwidth]{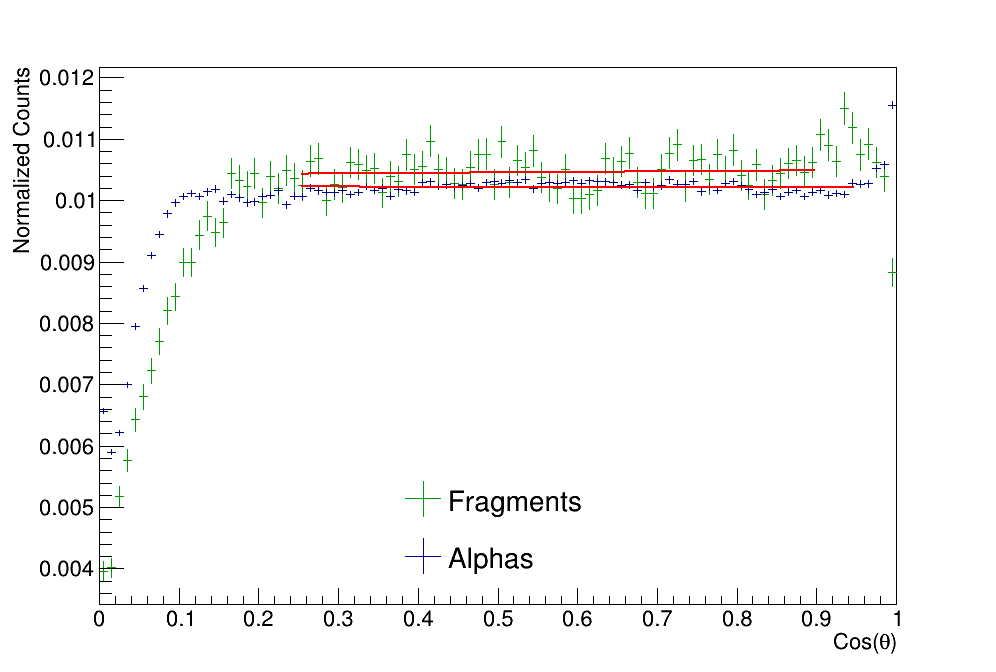}
		\caption{Area normalized fission fragment and alpha particle polar angle distributions from the $^{252}$Cf source after electron drift speed and electron diffusion corrections are applied.  The red lines are the fits demonstrating the reproduction of the isotropic emission of fragments and alphas.  }
		\label{fig:cf252polarAngle}
	\end{figure}

	Polar angle distributions for the fission fragments and alpha particles from the $^{252}$Cf source after the electron drift speed and electron diffusion corrections can be seen in Figure \ref{fig:cf252polarAngle}.  This plot demonstrates the fissionTPC's ability to reconstruct the polar angles of fission fragments over an angular range from about $0.25 < \cos\theta < 0.9$.  
	
	The drop off at low $\cos\theta$ is due to the fragments and alpha particles losing energy and ranging out in the target material, and the bump at high $\cos\theta$ is due to the saturation of the readout electronics for pads that have a large number of electrons deposited on them in a short amount of time. All the charge from a fission fragment that is emitted perpendicular to the anode will be collected on a few number of pads, causing saturation.  Due to the differentiation step in the processing from pad signals to voxels of charge, no charge is assigned to those voxels after saturation occurs, effectively creating a hole in the center of the charge cloud.  Because the fitting of the charge clouds is weighted by the ADC value in each voxel, a hole in the center of the charge cloud skews the angle in such a way that the saturation creates the ``bump" seen at high $\cos\theta$ in Figure \ref{fig:cf252polarAngle} \cite{hensleThesis}.
	
	\section{Linear Momentum Transfer}
	% Put \label in argument of \section for cross-referencing
	%\section{\label{}}
	Before a measurement of the anisotropy can take place, a measurement of linear momentum transfer is needed to convert from the lab frame to the center-of-mass frame.  The amount of linear momentum that is transferred from the incident neutron to the target nucleus shows up in the opening angle of the fission fragments \cite{Sikkeland1962}.  By placing the actinide deposits on a thin carbon foil, both fragments can be detected in the fissionTPC, allowing fission fragment opening angle measurements.  Coincident fission fragment pairs are selected and the angle between the two fission fragments is calculated after the drift speed and electron diffusion corrections are applied.
	
	\begin{figure}[]
		\includegraphics[width=.5\textwidth]{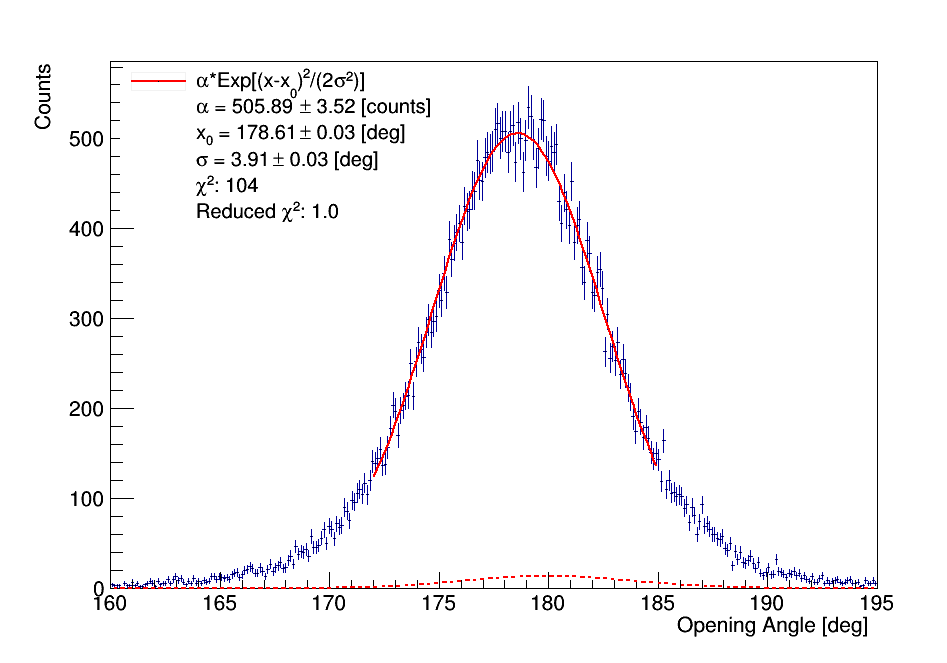}
		\caption{Fission fragment opening angles for 10 to 12 MeV incident neutrons on a $^{235}$U target.  Mean opening angles are taken from the solid line fit after the contribution from the dashed Gaussian that corrects for the wraparound neutrons is subtracted. }
		\label{fig:openingAngle}
	\end{figure}

\begin{figure}[]
		\includegraphics[width=.5\textwidth]{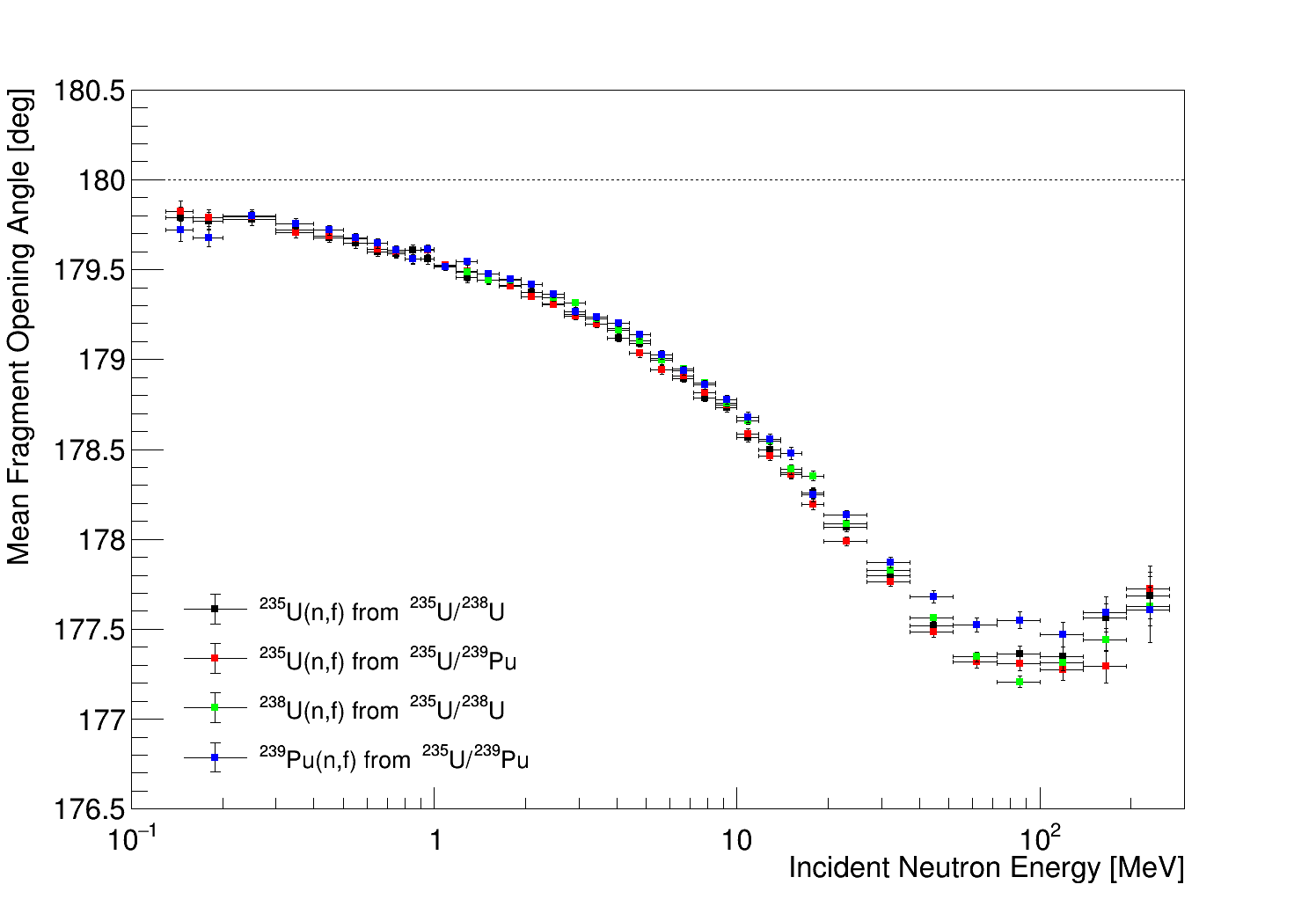}
		\caption{Measured fission fragment opening mean angles after wraparound subtraction. No momentum transfer would result in a mean opening angle of 180 degrees.}
		\label{fig:togetherFoldingAngle}
	\end{figure}

	Figure \ref{fig:openingAngle} shows the opening angle measurement between fission fragments that were produced by incident neutrons with energy between 10 and 12\,MeV, where the solid red line is a Gaussian fit to the data and the dotted red line is the wraparound Gaussian that is subtracted from the distribution to account for events from wraparound neutrons. The wraparound Gaussian distribution has an amplitude equal to the number of wraparound events in each energy bin as determined in Figure \ref{fig:nToF} and is centered at 180 degrees.  After performing the wraparound subtraction, the mean opening angle for each measured target and actinide can be plotted as a function of neutron energy, as shown in Figure \ref{fig:togetherFoldingAngle}.
	
	In order to extract the amount of linear momentum needed to reproduce these opening angles, a Monte Carlo simulation was used as the fissionTPC does not have the necessary information or precision to extract the linear momentum transfer on an event by event basis.
	
	A fission event is generated using the GEF code (Version 2017/1.1) \cite{GEFoverview} to get fission fragment masses and center of mass energies, and an initial downstream $\cos\theta$ is selected from the measured downstream $\cos\theta$ distribution.  Sampling from the measured angular distribution is necessary because the opening angle is dependent on both the amount of linear momentum transferred and the emission angle of the fragments.  An easy way to demonstrate this dependence is to consider fission fragments that are emitted along the  axis of the incident neutron beam -- no matter how much linear momentum is transferred to these fragments, the opening angle will always be 180 degrees in the lab frame.
	
	%\begin{figure}[]
	%	\includegraphics[width=.5\textwidth]{figures/simulationOverview.png}
	%	\caption{An outline of the simulation steps that are used to simulate a single fission fragment folding angle.  This process is repeated many times for each incident neutron energy bin and for a given neutron momentum transfer.}
	%	\label{fig:simulationOverview}
	%\end{figure}
	
	Next, that initial downstream $\cos\theta$ is converted from the lab angle to the center of mass angle via
	\begin{equation}
	\label{eqn:angleLabtoCMLMT}
	\cos\theta_{d,u}^{c.m.} = \sqrt{1 - \frac{E_{d,u}^{L}}{E_{d,u}^{c.m.}}(1 - \cos^{2}\theta_{d,u}^{L}) }.
	\end{equation}
	where the fragment energy in the lab frame can be calculated by \cite{hensleThesis}
	\begin{eqnarray}
	\label{eqn:ELabLMT}
	E_{d}^{L} = E_{d}^{c.m.} + \frac{m_{d} p_{n}^{2}}{m_{CN}^{2}}\left( \cos^{2}\theta_{d}^{L} - \frac{1}{2} \right) + \nonumber\\ \frac{m_{n} p_{n} \cos\theta^{L}}{m_{CN}^{2}} \sqrt{ \frac{2 E_{d}^{c.m.} m_{CN}^{2}}{m_{d}} + p_{n}^{2} \left( \cos^{2}\theta_{d}^{L} - 1\right)  }
	\end{eqnarray}
	for a given neutron momentum transfer $p_{n}$, mass of the compound nucleus $m_{CN}$, and downstream center of mass fragment energy $E_{d}^{c.m.}$ and mass $m_{d}$. $E_{d}^{c.m.}$ is computed according to the total kinetic energy value and masses sampled from the GEF output.  $m_{CN} = A + 1$ and the upstream mass is $m_{u} = m_{CN} - m_{d}$.  Treating fission as a two-body decay, in order to conserve momentum, the upstream center-of-mass angle must be the same as the downstream center of mass angle, i.e. $\cos\theta_{d}^{c.m.} = \cos\theta_{u}^{c.m.}$.  Converting this upstream center-of-mass angle back to the lab frame is done by inverting Equations \ref{eqn:angleLabtoCMLMT} and \ref{eqn:ELabLMT} using the appropriate fragment masses and energies \cite{hensleThesis}. The final step is to then take the opening angle between these upstream and downstream fragments and add a Gaussian smearing with a sigma of 3.8 degrees to match the angular resolution of the data.
	
	\begin{figure}[]
		\includegraphics[width=.5\textwidth]{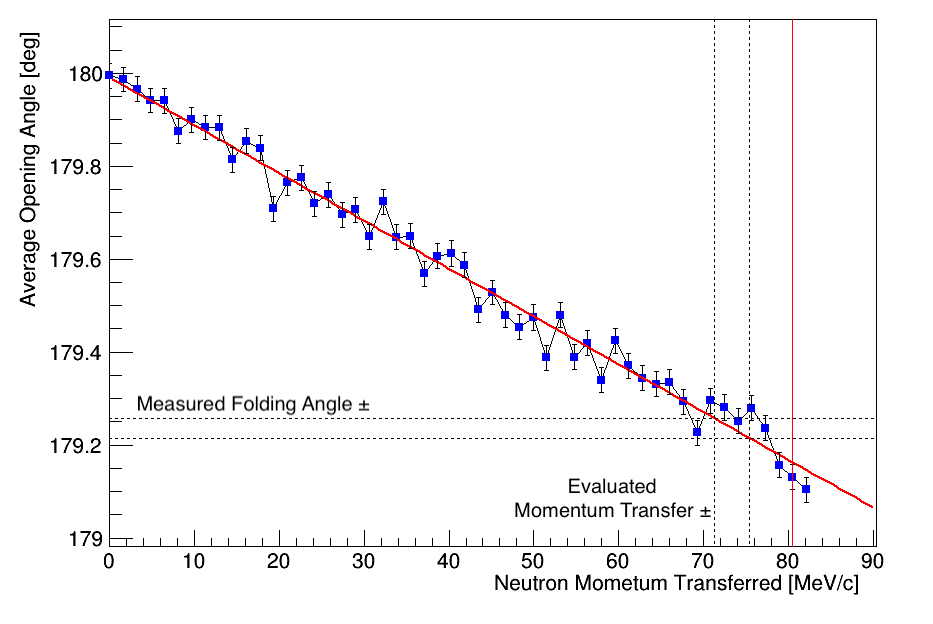}
		\caption{Simulated average opening angle for 3.4\,MeV incident neutrons as a function of applied momentum transfer.  See the text for more details.}
		\label{fig:simOutput}
	\end{figure}
	
	Repeating this process produces a set of opening angles for a particular incident neutron energy bin and neutron momentum transfer $p_{n}$.  The amount of neutron momentum transfer can be extracted by plotting the mean opening angles as a function of neutron momentum transfer and matching the measured opening angle to the simulation.  Figure \ref{fig:simOutput} demonstrates this procedure for a particular incident neutron energy bin and shows that the mean opening angle changes linearly as a function of applied neutron momentum transfer.  The horizontal dashed lines are the statistical uncertainty bounds from the measurement of the opening angle (from Figure \ref{fig:togetherFoldingAngle}), and the vertical dashed lines are where the opening angle measurement intersects with the linear fit in red.  Complete linear momentum transfer for this particular neutron energy bin is denoted by the solid vertical red line.
	
	\begin{figure}[]
		\includegraphics[width=.5\textwidth]{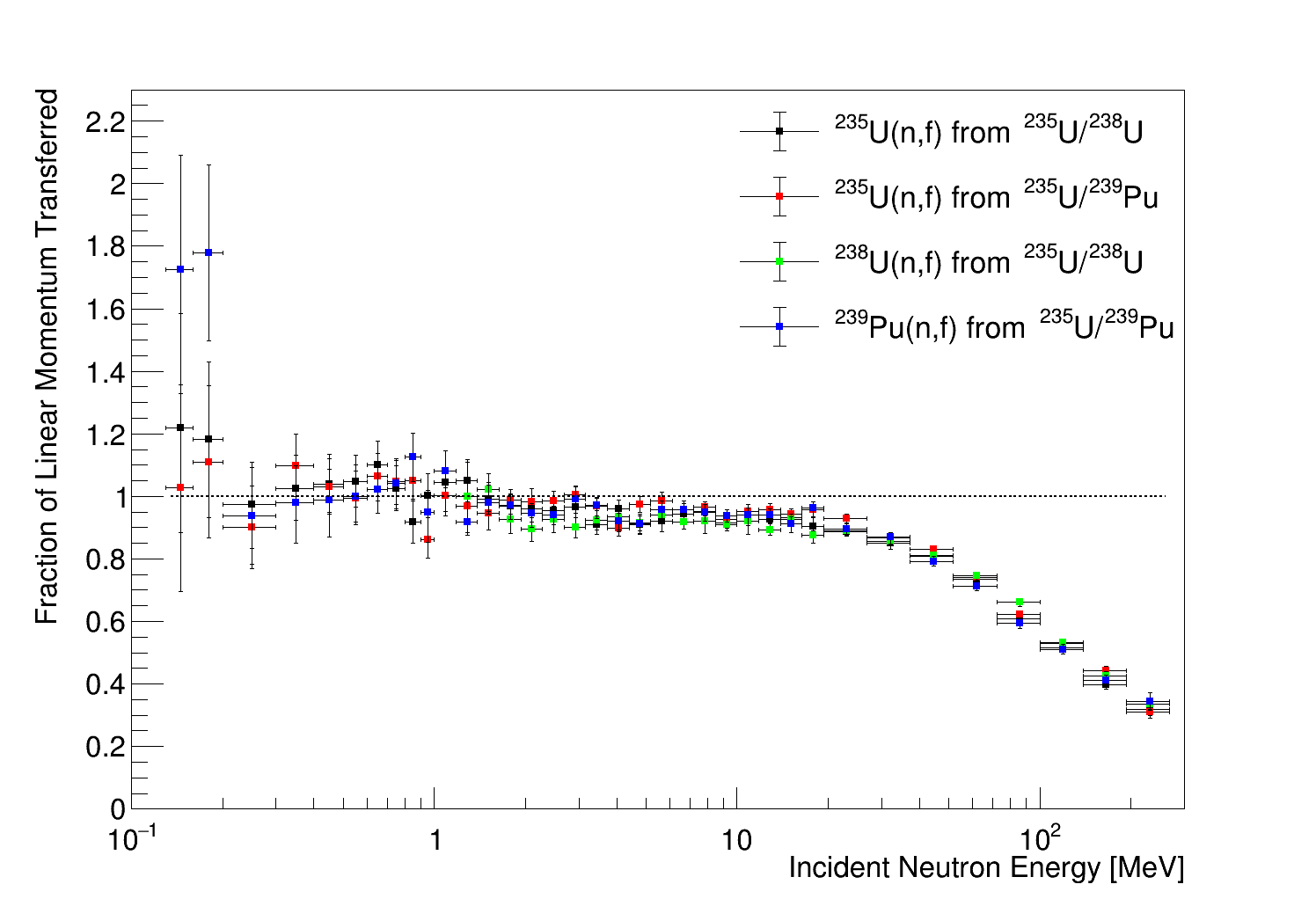}
		\caption{Fraction of the total momentum transfer as measured by the fissionTPC.  A normalization procedure was applied in order to achieve a weighted average of full momentum transfer for all neutron energy bins below 2\,MeV.}
		\label{fig:lmtfraction}
	\end{figure}
	
	Thus, in using this Monte Carlo simulation, the amount of linear momentum transferred from the incident neutron to the fissioning nucleus can be extracted; however, one additional normalization step is needed.  In order to reproduce full momentum transfer at low incident neutron energies, the electron drift speed difference applied between the upstream and downstream volumes is modified to create a less than 0.2 degree shift in the measured opening angles such that the weighted average is one in Figure \ref{fig:lmtfraction} for all incident neutron energy bins below 2\,MeV.  This normalization procedure accounts for the uncertainties in the electron drift speed and electron diffusion corrections between the upstream and downstream volumes \cite{hensleThesis}.
	
	Since $^{238}$U has a fission threshold of about 1.2\,MeV, the photofission peak from this actinide is not contaminated by wraparound neutrons.  After performing the complete normalization procedure, measuring the opening angle from these photon induced fission events gives $179.98 \pm 0.14$ degrees, which is consistent with the negligible momentum transfer expected from photons.
	
\begin{figure}[]
		\includegraphics[width=.5\textwidth]{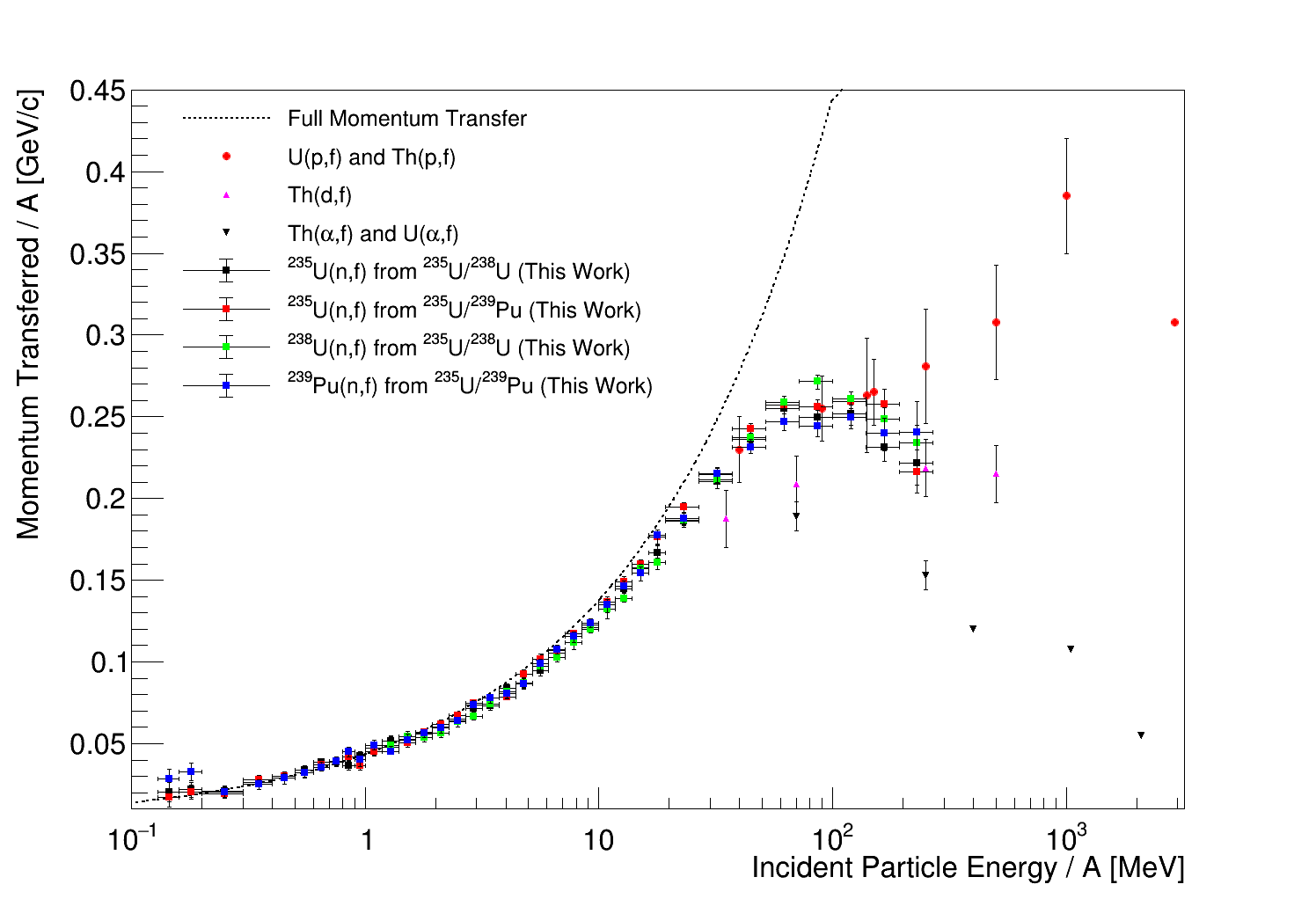}
		\caption{Average neutron linear momentum transfer as measured by the fissionTPC along with published measurements of linear momentum transfer for proton \cite{Meyer1979,SaintLaurent1982,Fatyga1985,Remsberg1969}, deuteron \cite{SaintLaurent1982}, and alpha \cite{Meyer1979,SaintLaurent1982} induced fission. }
		\label{fig:lmt}
	\end{figure}

	The final results of average linear momentum transfer as a function of incident neutron energy after the normalization procedure are shown in Figure \ref{fig:lmt}.
	
	\subsection{Uncertainty Analysis}
\begin{figure}[]
		\includegraphics[width=.5\textwidth]{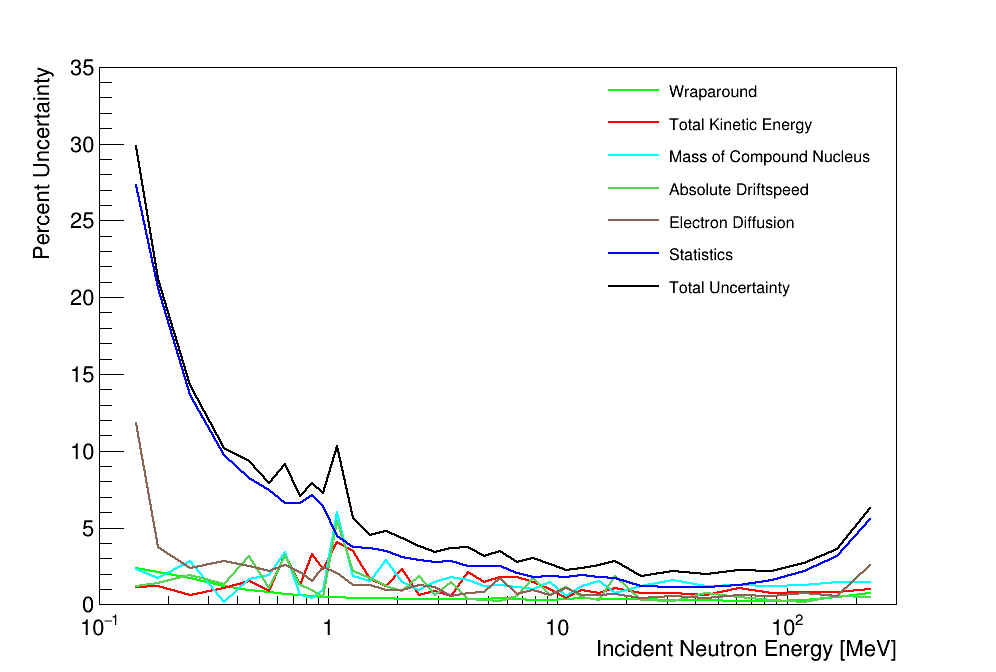}
		\caption{Uncertainty budget on the linear momentum transfer result for each systematic variation considered for the $^{235}$U half of the $^{235}$U/$^{238}$U target.  }
		\label{fig:lmterror}
	\end{figure}

	The error bars shown in Figures \ref{fig:lmtfraction} and \ref{fig:lmt} include both statistical and systematic uncertainties.  An example uncertainty budget is shown in Figure \ref{fig:lmterror} for the  $^{235}$U half of the $^{235}$U/$^{238}$U target measurement. 
	
	Statistical uncertainties from the fit of the measured opening angle are displayed in Figure \ref{fig:togetherFoldingAngle}.  The magnitude of these statistical uncertainties propagated through the simulation can be seen directly in Figure \ref{fig:simOutput} where the range of the opening angle measurement is equated to a range in the evaluated momentum transfer.  Statistical uncertainty is the dominant uncertainty, particularly at low incident neutron energies where the opening angle measurement requires fractions of a degree resolution.
	
	Many systematic uncertainties were also considered and a sensitivity study was performed for each of these to estimate their potential magnitude.  
	
	The contributions due to the neutron beam wraparound uncertainty is found by subtracting a different amplitude for the wraparound Gaussian shown in Figure \ref{fig:openingAngle} in accordance with the wraparound error and propagating the modified result through the simulation results.  The uncertainty on the percentage of neutron beam wraparound in each energy bin is calculated from the uncertainty on the fit that is integrated to get the neutron beam wraparound percentage. $^{238}$U has no need for the wraparound correction because of its roughly 1.2\,MeV fission threshold. Thus, the agreement between $^{238}$U and $^{235}$U/$^{239}$Pu provides validation that the neutron beam wraparound is handled correctly.
	
	Opening angle measurements and propagation through the Monte Carlo simulation were redone using different electron drift speed and electron diffusion values.  Electron drift speeds were varied by $ \pm 0.03$\,cm/\textmu s and the electron diffusion correction was varied by $\pm 10 \%$.  These variations were applied to both the upstream and downstream volumes equally.  The drift speed and diffusion difference between upstream and downstream volumes are taken into account using the normalization procedure.
	
	Total kinetic energy (TKE) of the two fission fragments and the mass of the fissioning nucleus are inputs into the Monte Carlo simulation used to extract the linear momentum transfer from the measurement of the opening angle.  Both of these parameters were varied and the entire simulation was redone in order to assess the sensitivity with respect to these input parameters.  Separate simulations were done for TKE values varied by 2\,MeV with respect to the mean value from the GEF code, and the mass of the compound nucleus was decreased by 5\,amu, showing up in either the downstream or upstream fragment -- also taking into account uncertainties in the mass distribution.  Taking mass from just the downstream or upstream fragment has a larger overall effect on the simulated opening angle than just moving the light or heavy mass peak from GEF.  Since upstream and downstream fragments have equal probability of being the light or heavy fragment, just moving the center of the mass distribution from GEF would average out much more than taking the mass exclusively from the upstream or downstream fragment.
	
	For all of these considered systematic variations, an independent linear momentum transfer result was computed. The uncertainty associated with each parameter variation was taken as the difference between the result when varying that parameter and the primary momentum transfer result.   These uncertainties were then treated independently and added in quadrature to get a total uncertainty, which includes both the statistical and systematic uncertainties.  Some parameters not having a linear relationship to the momentum transfer or an uneven parameter range (taking 5\,amu from the compound nucleus for example), produces slightly asymmetric error bars. The largest of the asymmetric uncertainty contributions for each incident neutron bin are shown in Figure \ref{fig:lmterror}.

	\subsection{Discussion}
	A few other sources of uncertainty were considered and warrant discussion but are not included in the error bars.  Most notably, fission is assumed to be a two body decay in the Monte Carlo simulation, but this ignores ternary fission.  Any ternary particle will certainly have an effect on the opening angle for that particular fission event, but we have found no published evidence that ternary particles have anisotropic emission with respect to the incident neutron beam direction.  In the context of this analysis, this means that there is no overall shift of the center of the opening angle distribution.   Additionally, ternary fission consists of less than 1\% of all fission events \cite{VandenboschText}, so any potential effect from an angular anisotropy with respect to the neutron beam would be minimal.
	
	Prompt neutron emission from the fragments can produce slight changes in the fragment angle.  In the rest frame of the fragment, the angular distribution of the emitted neutrons is measured to be primarily isotropic.  A small anisotropic component dependent on the angular momentum state of the fission fragment might exist, but is debated \cite{Chietera2010}; therefore, the uncertainty from prompt neutron emission is treated as negligible.
	
	One difference between the detection of fission fragments in the different detector volumes is that one of the fragments must travel through the carbon backing.  In order to ensure no angular bias is introduced in the fragment traveling through the backing, the two targets used in this analysis were placed in different orientations.  The $^{235}$U/$^{238}$U target has the actinides in the downstream volume, but the $^{235}$U/$^{239}$Pu target has the actinides in the upstream volume.  However, the Monte Carlo simulation samples from the downstream $\cos\theta$ distribution for both targets.  Since there is good agreement between these two targets, no detectable bias is introduced due to the fragments traveling through the carbon backing.
	
	To the best of our knowledge, this is the first measurement of neutron induced fission momentum transfer, so the comparison with other data shown in Figure \ref{fig:lmt} is with respect to proton, deuteron, and alpha induced fission.
	
	\section{Fission Fragment Anisotropy}
	Measurements of neutron-induced fission fragment angular anisotropy are also presented here for $^{238}$U and $^{235}$U.  By applying the electron drift speed and electron diffusion corrections, the isotropic distribution of alpha particles and fission fragments were successfully reproduced for a specific angular range, demonstrated in Figure \ref{fig:cf252polarAngle}.  No efficiency correction is therefore needed in the angular measurements of $^{238}$U and $^{235}$U angular anisotropy.
	
\begin{figure}[]
		\includegraphics[width=.47\textwidth]{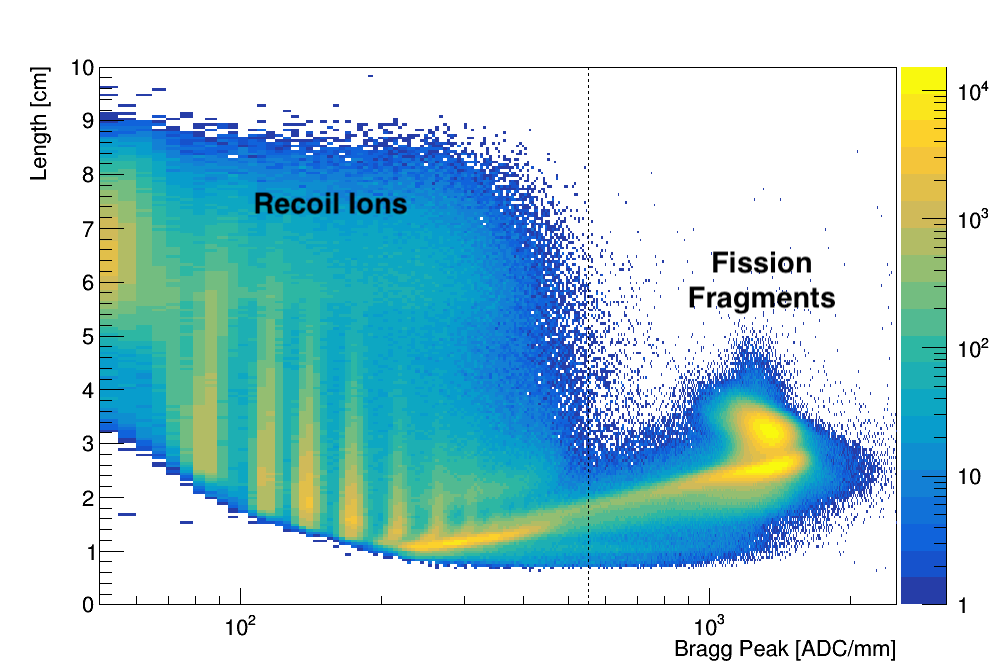}
		\caption{Track length versus Bragg peak showing separation of fission fragments from neutron recoils on detector materials and gas particles.  The vertical dashed line shows the location of the fragment selection cut.}
		\label{fig:lvbragg}
	\end{figure}

	The first step in the anisotropy measurement consists of separating fission fragments from recoil ions.  This is done by introducing a cut on the Bragg peak value, which provides more separation between fission fragments and argon recoils compared to a selection on total energy.  Track length versus Bragg peak is shown in Figure \ref{fig:lvbragg} as well as the location of the cut for this data from the $^{235}$U/$^{235}$U aluminum backed target.
	
\begin{figure}[]
		\includegraphics[width=.5\textwidth]{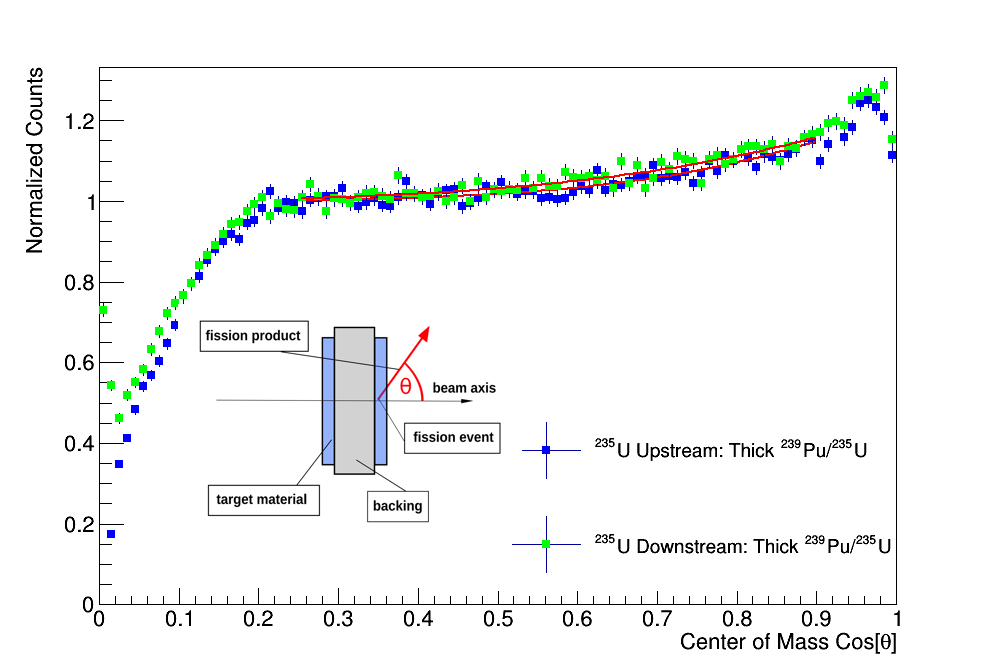}
		\caption{Example $^{235}$U angular distributions for 6 to 7 MeV neutrons from upstream and downstream measurements of the thick aluminum backed $^{239}$Pu/$^{235}$U target.  The angular distributions are normalized such that the fit equals one if extended to $\cos\theta = 0$.}
		\label{fig:angularDist}
	\end{figure}

	After selecting the fission fragments, they are separated into incident neutron energy bins and the electron drift speed and electron diffusion corrections are applied. Their angles are then converted from the lab frame to the center of mass frame using the average momentum transfer as measured in Figure \ref{fig:lmt}.  Even order Legendre polynomials up to fourth order are then used to fit the angular distributions.  An example upstream and downstream angular distribution is shown in Figure \ref{fig:angularDist} from the $^{235}$U side of the thick $^{235}$U/$^{239}$Pu target that was rotated with respect to the neutron beam.
	
	Extracting the measured anisotropy parameter from the Legendre polynomials is done via
	\begin{equation}\label{eqn:anisotropy}
	A _{\text{meas}} = \frac{W(\cos\theta_{c.m.} = 1)}{ W(\cos\theta_{c.m.} = 0)} = \frac{a_{0} + a_{2} + a_{4}}{a_{0} - a_{2}/2 + 3a_{4}/8}
	\end{equation}
	where $a_{n}$ corresponds to the coefficient of the n-th Legendre polynomial.  However, this measurement includes fission events from wraparound neutrons that do not belong in this incident neutron energy bin.  The measured anisotropy is a combination of the real anisotropy and the anisotropy of the wraparound neutron events, or more explicitly,
	\begin{equation}
	A_{\text{meas}} = (1-p_{\text{wrap}})A_{\text{real}} + p_{\text{wrap}} A_{\text{wrap}}
	\end{equation}
	where $p_{wrap}$ is the percentage of wraparound neutrons in that incident neutron energy bin, as plotted in Figure \ref{fig:nToF}.  With the assumption that the low energy wraparound neutrons produce an isotropic distribution of fission fragments, $A_{wrap} = 1$ and the real value of the anisotropy is 
	\begin{equation}
	\label{eqn:Areal}
	A_{\text{real}} = \frac{A_{\text{meas}} - p_{\text{wrap}}}{1-p_{\text{wrap}}}.
	\end{equation}
	
\begin{figure}
		\includegraphics[width=0.5\textwidth]{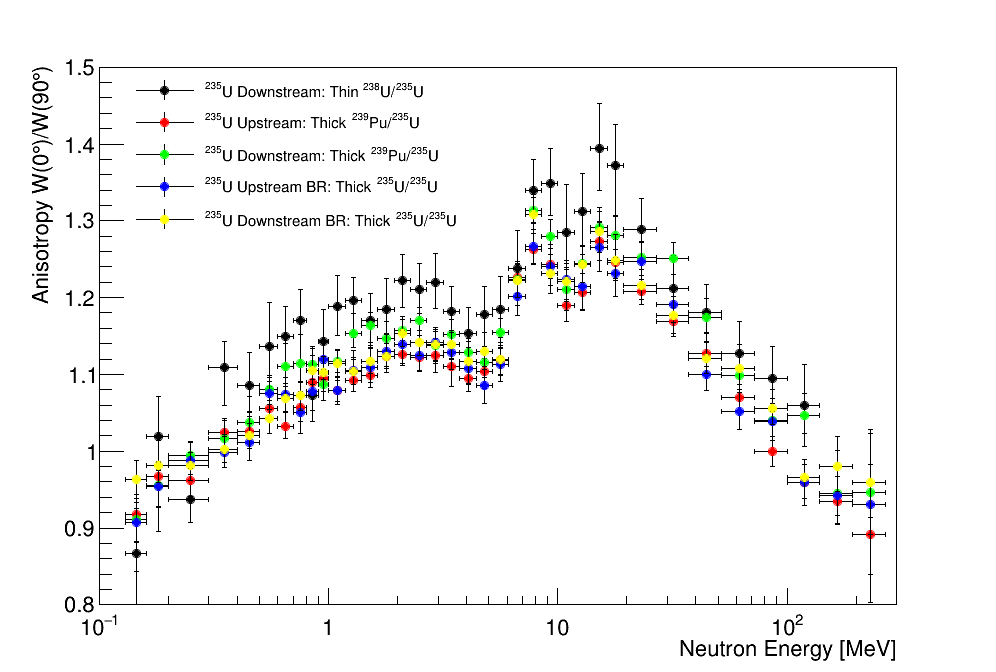}
		\caption{Anisotropy parameter, as defined by Eqn. \ref{eqn:A}, from each target and orientation of $^{235}$U. }
		\label{fig:u235anisotropyAll}
	\end{figure}

\begin{figure}
		\includegraphics[width=0.5\textwidth]{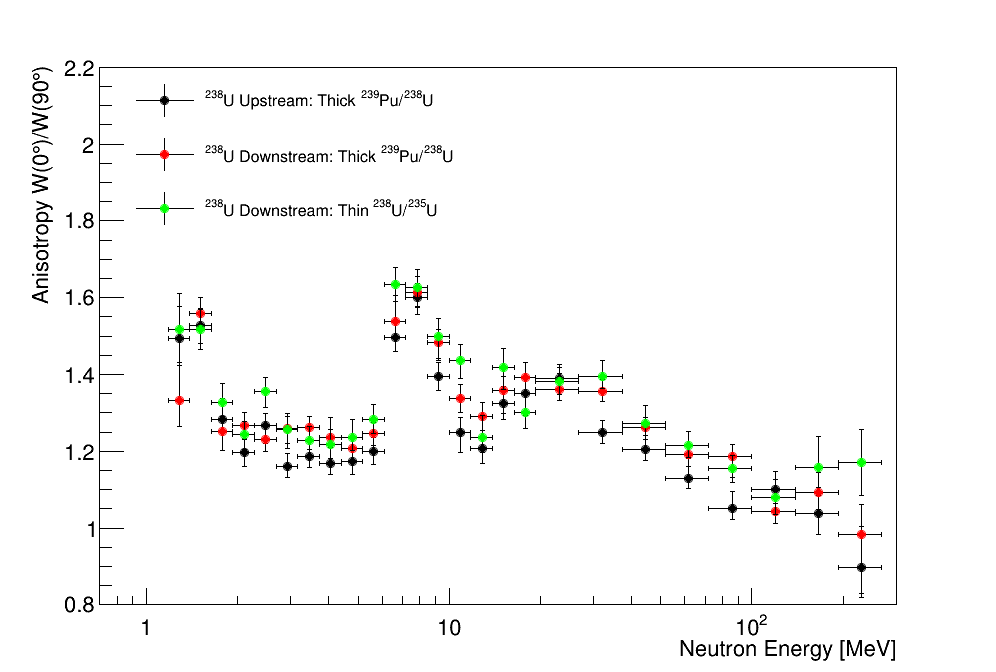}
		\caption{Anisotropy parameter, as defined by Eqn. \ref{eqn:A}, from each target and orientation of $^{238}$U. }
		\label{fig:u238anisotropyAll}
	\end{figure}

	Each of the different actinide targets depicted in Figure \ref{fig:targets} were placed in the fissionTPC to measure the anisotropy (with the exception of the thin-backed $^{235}$U/$^{239}$Pu target).  Each measurement had slightly different operating parameters resulting in slightly different corrections for electron drift speed and electron diffusion.  Measuring the anisotropy with many different targets demonstrates the reproducibility of the fissionTPC over the course of the experiments.  $^{235}$U and $^{238}$U anisotropy results ($A_{real}$) can be seen in Figures \ref{fig:u235anisotropyAll} and \ref{fig:u238anisotropyAll} for each of the targets.
	
	\subsection{Uncertainty Analysis}
	The uncertainty analysis procedure for the anisotropy results is very similar to the procedure followed for the linear momentum transfer results -- parameters are varied and the difference between the primary anisotropy result and that variation is taken as the uncertainty associated with that parameter.
	
	Statistical uncertainties on the anisotropy analysis are propagated from Equation \ref{eqn:anisotropy} where the uncertainties on $a_{n}$ are taken directly from the fit.  Similarly, uncertainty from neutron beam wraparound is propagated directly from Equation \ref{eqn:Areal} using the uncertainty on the percent of the wraparound counts in each incident neutron energy bin.  
	
	Electron drift speed values are varied by $\pm 0.01$\,cm/\textmu s and the electron diffusion correction is varied by $\pm 10\%$ based on the measurement techniques laid out in Sections \ref{sub:electronDrift} and \ref{sub:electronDiff}.  Modifying the drift speed and diffusion are roughly degenerate with the second order Legendre polynomial, so the uncertainties on the drift speed measurement and diffusion correction are essentially constant across all incident neutron energy bins at about 1\% and 0.5\% of the anisotropy, respectively.
	
	\begin{figure}
		\includegraphics[width=0.5\textwidth]{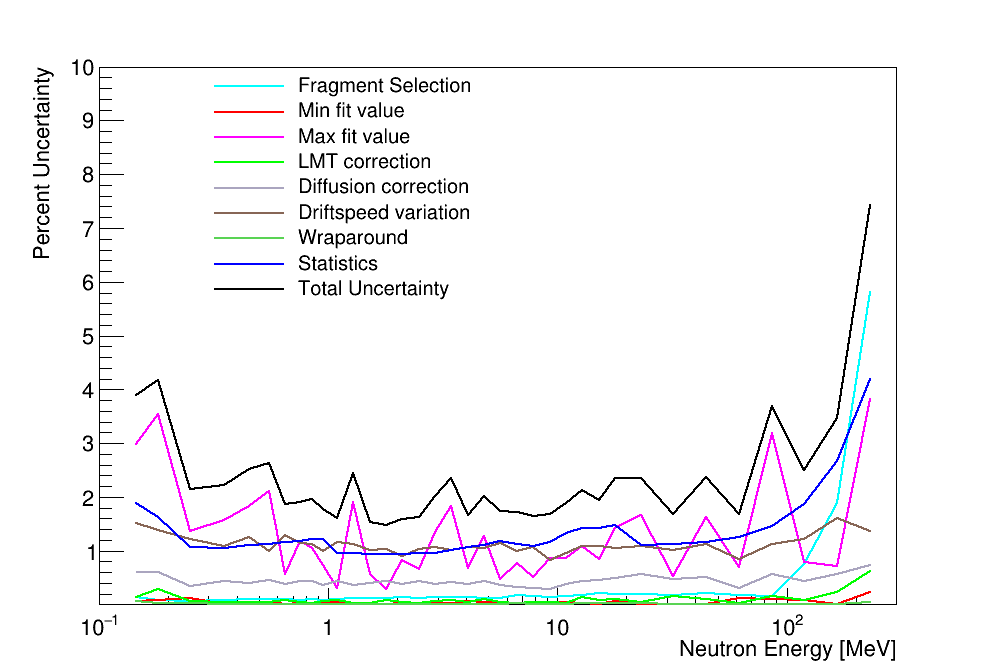}
		\caption{Uncertainty budget for $^{235}$U anisotropy from the thick backed $^{235}$U/$^{239}$Pu target. }
		\label{fig:u235anisotropyError}
	\end{figure}
	
	Lower and upper bounds of the Legendre polynomial fits are also varied to acquire an uncertainty associated with the fit range.  Less than a 0.02\% change in the anisotropy result is observed when varying the fit lower bound by $\pm 0.03 \cos\theta$, but when varying the fit upper bound by that same amount, large differences of up to 4\% arise. When changing the upper bound of the fit, the fit is affected by the bump at large $\cos\theta$ from the detector saturation effect discussed in Section \ref{sub:cf252}.  This is compounded by the fourth order Legendre component also having the most influence at larger $\cos\theta$.  Due to these compounding factors, the fit upper bound is actually the largest uncertainty contribution for the majority of incident neutron energy bins.
	
	The location of the Bragg peak cut used to select the fission fragments shown in Figure \ref{fig:lvbragg} is varied by $\pm 50$\,ADC/mm.  At very high incident neutron energies, this cut starts to dip down into the high energy recoil ions which show up at large $\cos\theta$.  These high energy recoils lead to an increased uncertainty associated with the fission fragment selection, which can be seen in Figure \ref{fig:lmterror}.
	
	The kinematic correction to convert from lab to center of mass angles is varied in accordance with the uncertainty from the linear momentum transfer result in Figure \ref{fig:lmt}.
	
	\subsection{Anisotropy Results and Discussion}
	In order to report a single anisotropy result from the many fissionTPC data sets seen in Figures \ref{fig:u235anisotropyAll} and \ref{fig:u238anisotropyAll}, each measurement was combined with a weighted average
	\begin{equation}
	\bar{A} = \frac{\sum w_{i}A_{i}}{\sum w_{i}}
	\end{equation}
	where the squared weights are given by the inverse average of the slightly asymmetric error bars for each data point
	\begin{equation}
	w_{i} = \left(\frac{2}{\sigma_{i}^{+} + \sigma_{i}^{-}}\right)^{2}.
	\end{equation}
	The uncertainty reported on this weighted average is the greater of the standard uncertainty expression for the weighted average
	\begin{equation}
	\label{eqn:wavgerr}
	\sigma_{\bar{A}} = \frac{1}{\sqrt{\sum w_{i}}}
	\end{equation} 
	or the weighted standard deviation
	\begin{equation}
	\label{eqn:wstd}
	\sigma_{\bar{A}} = \sqrt{ \frac{\sum w_{i}(A_{i} - \bar{A})^{2}}{N^{-1}(N - 1)\sum w_{i}} }
	\end{equation}
	where N is the number of data points with non-zero weight.  The need to compare two different uncertainties arises from the limitations of each method. The typical weighted average uncertainty (Eq. \ref{eqn:wavgerr}) does not include the spread of the data points and the weighted standard deviation (Eq. \ref{eqn:wstd}) produces an uncertainty of zero if all points lie on top of each other, regardless of the size of the error bars.  Taking the largest uncertainty between these two methods ensures that both the spread of the data and size of the error bars are taken into account.

	\begin{figure}
		\includegraphics[width=0.5\textwidth]{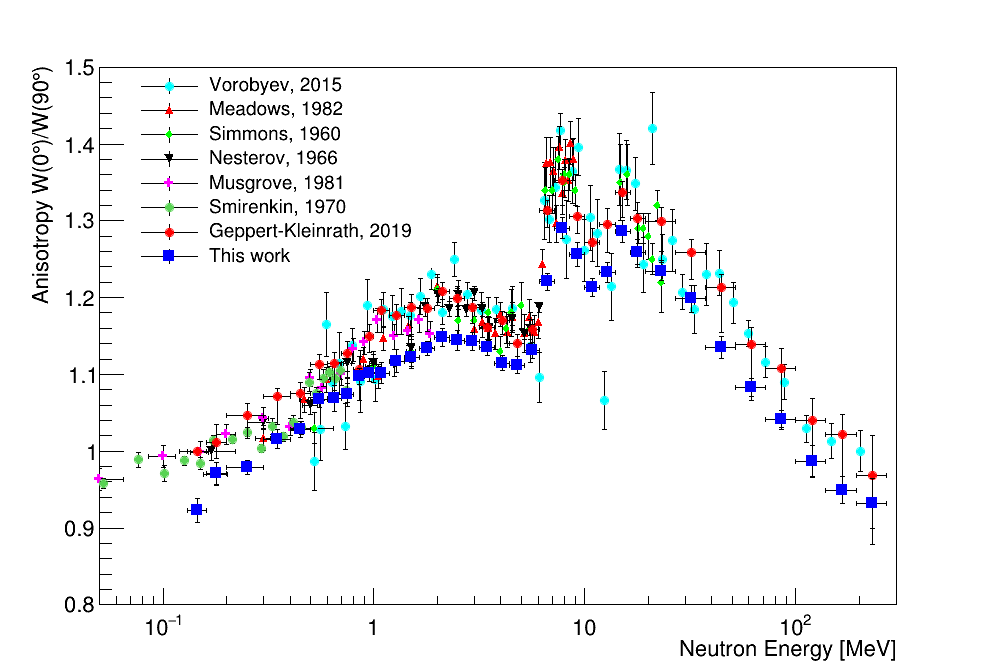}
		\caption{Anisotropy results combining the $^{235}$U data sets compared to available data in EXFOR \cite{Otuka2014}. }
		\label{fig:u235anisotropyResult}
	\end{figure}

	\begin{figure}
		\includegraphics[width=0.5\textwidth]{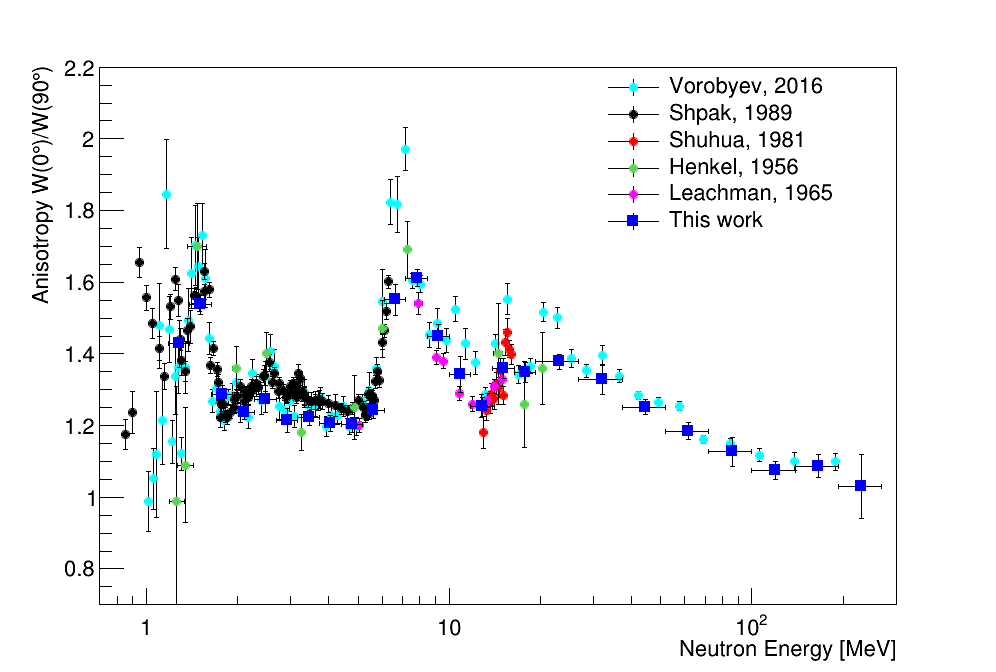}
		\caption{Anisotropy results combining the $^{238}$U data sets compared to available data in EXFOR \cite{Otuka2014}. }
		\label{fig:u238anisotropyResult}
	\end{figure}
	
	Comparison of the anisotropy results to the other published data from EXFOR \cite{Otuka2014} shows generally good agreement for $^{238}$U, but not $^{235}$U.  All of the previously published $^{235}$U data shown here use a normalization at low incident neutron energies in order to determine their detection efficiency.  This process was also used in a previous measurement of the $^{235}$U anisotropy by the fissionTPC, which shows good agreement with the published data \cite{Verena}.  However, with the additional electron drift speed and diffusion corrections applied, no normalization is needed for this work. 
	
	Better agreement with the $^{238}$U data can be explained by the lack of low incident neutron energies being used for detector response calibration and efficiency.  In particular, the EXFOR entry for the $^{235}$U Vorobyev data references a publication that assumes full detection efficiency for $0.4 < \cos\theta < 1.0$ \cite{Vorobyev2015}.  However, the $^{238}$U Vorobyev EXFOR entry points to a reference which shows subsequent calibration work with a $^{252}$Cf source \cite{Vorobyev2016} demonstrating incomplete efficiency for $0.4 < \cos\theta < 1.0$, thus possibly explaining why the $^{238}$U results show better agreement than $^{235}$U.

	\section{Summary}
	The fissionTPC is studying fission in a novel way with the use of its full three-dimensional tracking ability. This ability requires an extensive reconstruction process to transform raw signals into individual track parameters, including finding the electron drift speed by flattening the spontaneous alpha polar angle distribution as well as applying a correction to the fission fragment angles due to the electron diffusion coefficients not being equal in the drift and radial directions of the fissionTPC.  By applying this reconstruction process, the polar angle distribution of spontaneous fission and alpha emissions from a $^{252}$Cf source was successfully shown to be isotropic.
	
	Placing actinide deposits on a thin-carbon backing allows for the detection of both the upstream and downstream fission fragments in the fissionTPC.  By measuring the opening angle between the upstream and downstream fission fragments and using a Monte Carlo simulation, the first measurement of linear momentum transfer from incident neutrons to the fissioning nucleus was performed.
	
	This linear momentum transfer measurement was then applied in the conversion of fission fragment angles from the lab frame to the center of mass frame to determine the angular anisotropy of the fission fragments as a function of incident neutron energies.  Both the linear momentum transfer and anisotropy measurements included careful attention to systematics whose exploration was made possible by the wealth of data for every fission event detected in the fissionTPC. 
	
	Measurements of fission fragment angular anisotropy show significant disagreement from published $^{235}$U data. This might be explained via the detection efficiency procedures used in previous analyses which assumed isotropy at low incident neutron energies.  The $^{238}$U angular anisotropy presented here agrees better with previous publications where those detection efficiency procedures cannot be performed due to the fission threshold at roughly 1.2\,MeV.

	\begin{acknowledgments}
		This work was performed under the auspices of the U.S. Department of Energy by Lawrence Livermore National
		Laboratory under Contract No. DE-AC52-07NA27344. The neutron beam for this work was provided by LANSCE, which is funded by the U.S. Department of Energy and operated by Los Alamos National Security, LLC, under Contract No. DE-AC52-06NA25396. University collaborators acknowledge support for this work from the U.S. Department of Energy Nuclear Energy Research Initiative Project No. 08-014, the DOE-NNSA Stewardship Science Academic Alliances Program, under Grant No.  DE-NA0002921, and through subcontracts from LLNL and LANL.
	\end{acknowledgments}
	
	% Create the reference section using BibTeX:
	\bibliography{references}
	
\end{document}